# Solver-consistent learned curvature for algebraic volume-of-fluid simulations of rising bubbles: symmetry-preserving networks, closed-loop data aggregation and a posteriori benchmarks

Parisa Tayerani and Mahmood Mousavi

School of Engineering, Embry-Riddle Aeronautical University, Daytona Beach, FL 32114, USA

Corresponding author: mousavs1@erau.edu

## Abstract

Curvature reconstructed from compressed volume fractions exhibits a persistent force-weighted bias in the tested algebraic volume-of-fluid formulation and motivates a solver-consistent learned correction. In OpenFOAM's `interFoam`, face curvature underestimates the exact value by 11–22 % for radii of 2.5–40 cells. This bias explains the observed 14–17 % static Laplace-pressure deficit and is consistent with approximately 2 % rise-velocity underprediction in the first Hysing benchmark. We estimate curvature from 5×5 volume-fraction stencils using a two-hidden-layer perceptron with exact phase-exchange antisymmetry and invariance under the eight square-grid symmetries. Training uses solver-generated profiles from periodic translating bubbles, droplets, and flat films with exact analytical reference curvatures to match the deployed interface representation. Iterative data aggregation incorporates model–solver feedback and resolves deployment instabilities observed in 15 of the 76 periodic cases evaluated with the initial model (19.7 %). Band-weighted face interpolation integrates the estimator into the existing pressure–velocity coupling, with the same capillary face force entering momentum prediction and pressure correction. On independent evaluation configurations, learned curvature reduces static pressure-jump errors to 2–6 % and time-averaged maximum capillary numbers by factors of 2.8–16. Across three grids, time-integrated rise-velocity errors decrease from 2.06–2.76 % to 0.65–1.04 % in the first benchmark and from 5.64–8.72 % to 4.72–7.61 % in the second. Tuned curvature smoothing produces smaller static pressure-jump errors, but its first-benchmark accuracy deteriorates with refinement and its second-benchmark gains remain marginal. A separate interface-gated, dissipative momentum closure achieves held-out $R^2$=0.47 on Favre-filtered targets, yet changes rise velocity by less than $3\times10^{-4}$ of its peak value. These comparisons identify capillary-force discretisation as the more influential correction target for the tested laminar bubbles.

## 1 Introduction

Rising bubbles test multiphase solvers through coupled interface deformation, material-property discontinuities, and concentrated capillary forces, with shapes ranging from compact interfaces to skirted structures and satellite bubbles [1, 2]. Complex rheology introduces further coupling between material response and interfacial dynamics [3], while interface-resolved simulations determine the surrounding flow without empirical bubble drag or lift laws. Established interface representations include front tracking [4], level sets [5, 6], and diffuse-interface formulations that couple fluid motion to an evolving phase field [7–14]. The volume-of-fluid (VOF) method conservatively transports cell-averaged phase fractions [15] through conservative fluxes and bounded advection that preserve phase volume and admissible composition. Algebraic methods use compressive differencing [16] or flux-corrected transport [17], whereas geometric methods reconstruct interfaces and calculate fluxes from cell-scale geometry [18–20]. We examine the algebraic formulation in OpenFOAM's `interFoam` solver [21–23] using the Hysing rising-bubble benchmarks [24]. The original three groups compared centroid height, rise velocity, and circularity using finite-element level-set and arbitrary-Lagrangian–Eulerian methods. Their case-1 results agreed to approximately three to four significant digits across these independently developed numerical approaches. Case 2 combines a density ratio of 1000 with weaker surface tension and produces skirts and trailing filaments whose breakup increases method-dependent variation. Subsequent diffuse-interface [7], finite-element level-set [6], geometric VOF [25], and unequal-density Navier–Stokes–Cahn–Hilliard studies [8] establish complementary comparisons for these configurations. Their use in structure-preserving and energy-stable formulations also supports examining curvature-related errors within a common benchmark setting [9–14].

The continuum-surface-force (CSF) model expresses the interfacial stress jump as a volumetric force determined by curvature and the phase-indicator gradient [26]. Balanced-force discretisation combines compatible capillary and pressure-gradient operators [27, 28] to reduce numerical imbalances that generate parasitic currents. The origins and scaling of these non-physical interfacial velocities are well documented [29, 30]. Curvature obtained from smoothed or compressed volume fractions need not converge under refinement, which motivates height-function and reconstruction-based geometric estimators [28, 31, 32]. Height-function convergence also depends on interpolation: constant column-wise estimates give first-order accuracy, whereas linear or quadratic interpolation recovers second-order accuracy in the analysed configurations [33]. Static and oscillating bubbles provide complementary tests of direct-curvature, Laplace–Beltrami, and curvature-evolution formulations [34], including equilibrium and dynamic capillary response. For algebraic VOF, geometric estimators must accommodate profiles shaped by compression and flux limiting, while conventional `interFoam` differentiates the normalised volume-fraction gradient directly. Explicit surface-tension forcing additionally imposes a capillary-wave time-step restriction proportional to $h^{3/2}$ [26, 35], even when bulk velocities are negligible. Artificial interfacial viscosity offers a complementary route for damping spurious capillary modes through local dissipation in unstructured VOF solvers [36].

Machine learning supplies flow surrogates, constitutive closures, and learned numerical components [37–42], including physics-informed networks [43] and operator models for interface evolution and bubble dynamics [44–46]. Multiphase applications span large datasets [47], transformer-based boiling prediction [48], porous-media displacement [49], video-based droplet reconstruction [50], particle-graph free-surface models [51], and Kolmogorov–Arnold networks [52, 53]. Learned closures retain the conservative solver and extend subgrid modelling [54–56] through invariance [57, 58], realisability [59], scaling constraints [60], or compact functional representations [61]. Explicit filtering addresses training–deployment consistency [62], while structure-preserving architectures enforce energy conservation independently of network weights [63, 64]; constrained lattice-Boltzmann closures provide another example [65]. These developments emphasise coordination among data generation, training, and deployment [66]. Differentiable solvers [67–70], unrolled optimisation [71, 72], and learned nudging [73] account for the coupled numerical response during model development. Uncertainty estimates and conservative guards help calibrate model confidence and constrain corrections when solver deployment extends beyond the training distribution [74–76]. OpenFOAM implementations support network deployment through embedded TensorFlow [77], SmartSim coupling [78], surrogate–solver switching [79], and learned pressure initialisation [80]. For two-phase applications, Favre-filtered equations additionally introduce unresolved interfacial contributions [81], while data-driven closures also target interfacial area [82]. Geometry-focused models retain the conservative solver and learn curvature from volume-fraction stencils [83, 84] through symmetry-preserving architectures or height-function inputs [85, 86]. Related geometric targets include three-dimensional interface normals [87], reconstruction on unstructured meshes [88], and local flux fractions for conservative phase transport [89]. Many curvature models use exact circle or sphere intersections [83–86], whereas algebraic transport generates two- or three-cell profiles dependent on compression, limiting, flow direction, Courant number, and parasitic motion. Predicted capillary forces subsequently modify these inputs and create the closed-loop covariate shift recognised in imitation learning [90] and constitutive modelling [41, 67, 71].

We therefore quantify conventional `interFoam` curvature bias and its relationship to discrete Laplace-pressure balance, then construct a curvature estimator using solver-generated volume-fraction stencils. The network preserves phase-exchange antisymmetry and Cartesian-grid invariance, while iterative data aggregation accounts for feedback between capillary predictions and subsequent interface profiles. Band-weighted face interpolation couples the estimator to the pressure–velocity formulation, while conventional curvature provides the reference path for controlled solver comparisons. A static bubble and both Hysing benchmarks test learned, conventional, and smoothed CSF on three grids, with rising-bubble configurations excluded from curvature training. The assessment combines curvature accuracy, capillary equilibrium, parasitic currents, bubble dynamics, and numerical conservation to distinguish mean geometric accuracy from coupled deployment behaviour. An interface-gated dissipative momentum closure trained on Favre-filtered fine-grid bubbles separately assesses unresolved momentum transport in the same laminar configurations. Section 2 presents the governing equations and constitutive closure, and Sections 3 and 4 describe learned curvature and its numerical implementation. Section 5 defines configurations and diagnostics, Section 6 examines the numerical results, and Section 7 summarises the principal findings.

# 2 Mathematical formulation

## 2.1 One-fluid model

We consider two incompressible, immiscible, isothermal Newtonian fluids occupying a fixed domain $\Omega\subset\mathrm{R}^d$, with $d\in\{2,3\}$, separated by a sharp interface $\Gamma(t)$. Subscripts 1 and 2 denote the liquid and gas, respectively, and the liquid volume fraction $\alpha(\mathbf{x},t)$ identifies the local phase distribution. The phase fraction equals one in pure liquid and zero in pure gas, with intermediate values within the numerically resolved interface band. Linear interpolation of the constant phase properties defines the mixture density and dynamic viscosity consistently with the transported liquid volume fraction:

$$\rho=\alpha\rho_1+(1-\alpha)\rho_2, \qquad \mu=\alpha\mu_1+(1-\alpha)\mu_2. \tag{1}$$

The mixture kinematic viscosity is $\nu=\mu/\rho$, and a single velocity field $\mathbf{u}$ satisfies incompressibility throughout both phases and the interfacial region:

$$\nabla\cdot\mathbf{u}=0. \tag{2}$$

The pressure $p$ enforces the divergence-free constraint, while the conservative momentum balance combines viscous stresses, gravitational acceleration, and capillary forcing within the one-fluid formulation:

$$\begin{aligned}\frac{\partial(\rho\mathbf{u})}{\partial t}+\nabla\cdot(\rho\mathbf{u}\otimes\mathbf{u})=-\nabla p+\nabla\cdot[2(\mu+\mu_{\mathrm{ML}})\mathbf{S}^d]\\ +\rho\mathbf{g}+\mathbf{f}_\sigma.\end{aligned} \tag{3}$$

Here, $\mathbf{g}$ denotes gravitational acceleration, $\mathbf{f}_\sigma$ is the capillary force per unit volume, and $\mu_{\mathrm{ML}}\geq 0$ is the learned dynamic viscosity introduced in Section 2.4. The viscous term uses the symmetric deviatoric strain-rate tensor and retains OpenFOAM's three-component trace convention for both three-dimensional and extruded two-dimensional calculations:

$$\mathbf{S}^d = \frac{1}{2}\left(\nabla\mathbf{u} + (\nabla\mathbf{u})^{\mathrm{T}}\right) - \frac{1}{3}(\nabla\cdot\mathbf{u})\mathbf{I}. \tag{4}$$

The rotation tensor $\mathbf{W} = \frac{1}{2}\left(\nabla\mathbf{u} - (\nabla\mathbf{u})^{\mathrm{T}}\right)$ defines the rotation rate $w = \sqrt{2\mathbf{W}:\mathbf{W}}$, while $s = \sqrt{2\mathbf{S}^d:\mathbf{S}^d}$ measures the strain rate used in the closure. Setting $\mu_{\mathrm{ML}} = 0$ and selecting conventional curvature recovers the laminar one-fluid formulation implemented in the OpenFOAM `interFoam` solver [22, 23].

## 2.2 Phase-fraction transport and interface compression

The algebraic VOF formulation advects the phase fraction with the divergence-free velocity and adds an artificial compressive flux to counteract numerical interface diffusion [21, 91]:

$$\frac{\partial\alpha}{\partial t} + \nabla\cdot(\alpha\mathbf{u}) + \nabla\cdot[\alpha(1-\alpha)\mathbf{u}_r] = 0. \tag{5}$$

The auxiliary velocity $\mathbf{u}_r$ supplies numerical compression rather than physical phase slip, and its face-based representation is specified through the compressive volumetric flux:

$$\phi_{r,f} = c_\alpha \frac{|\phi_f|}{|\mathbf{S}_f|}\, \hat{\mathbf{n}}_f\cdot\mathbf{S}_f. \tag{6}$$

The volumetric face flux is $\phi_f = \mathbf{u}_f\cdot\mathbf{S}_f$, where $\mathbf{S}_f$ is the outward face-area vector and $\hat{\mathbf{n}}_f$ denotes the interpolated interface normal. The compression coefficient is fixed at $c_\alpha = 1$, while $\alpha(1-\alpha)$ localises the artificial flux to the interface band and suppresses compression in pure phases. With periodic or impermeable boundaries, consistent discrete fluxes preserve phase volume, while compression and numerical diffusion determine the flow-dependent width and shape of the interface profile. Because compression is numerical, its contribution also enters the mass flux used by the discrete momentum equation. For constant phase densities, Eqs. (1) and (5) imply the following conservative density transport:

$$\begin{aligned} \mathbf{J}_r &= (\rho_1 - \rho_2)\alpha(1-\alpha)\mathbf{u}_r, \\ \frac{\partial\rho}{\partial t} &+ \nabla\cdot(\rho\mathbf{u} + \mathbf{J}_r) = 0. \end{aligned} \tag{7}$$

Equations (3) and (10) state the physical one-fluid balance. Before flux limiting, the corresponding numerical convective momentum flux has components $u_i\left(\rho u_j + J_{r,j}\right)$. The implementation uses the limited phase-flux counterpart for momentum convection, as specified in Section 4.

## 2.3 Surface tension and the modified pressure

The conventional formulation reconstructs the interface normal from the regularised volume-fraction gradient and computes signed curvature from its negative divergence for use in capillary forcing:

$$\hat{\mathbf{n}} = \frac{\nabla\alpha}{|\nabla\alpha| + \delta_N}, \qquad \kappa = -\nabla\cdot\hat{\mathbf{n}}. \tag{8}$$

The regularisation constant has inverse-length units and follows $\delta_N = 10^{-8}/\langle V\rangle^{1/3}$ in the OpenFOAM-7 reference implementation [22], where $\langle V\rangle$ is the mean cell volume. In two dimensions, a gas bubble of radius $R$ has $\kappa = -1/R$, whereas a liquid droplet has $\kappa = +1/R$ under the same phase-labelling convention. The CSF model supplies $\mathbf{f}_\sigma = \sigma\kappa\nabla\alpha$ for constant $\sigma$ and couples signed curvature directly to the transported phase-fraction gradient [26]. The conventional discrete operator first calculates the cell gradient, interpolates it to faces, and normalises the interpolated gradient before taking its finite-volume divergence:

$$\boldsymbol{\gamma}_P = (\nabla_h \alpha)_P, \qquad \boldsymbol{\gamma}_f = I_f(\boldsymbol{\gamma}_P),$$
$$\widehat{\mathbf{n}}_f = \frac{\boldsymbol{\gamma}_f}{|\boldsymbol{\gamma}_f| + \delta_N}, \tag{9}$$
$$\kappa_P = -\frac{1}{V_P} \sum_{f \in \partial P} \widehat{\mathbf{n}}_f \cdot \mathbf{S}_{Pf}.$$

Here, $I_f$ denotes cell-to-face interpolation, $V_P$ is the cell volume, and $\mathbf{S}_{Pf}$ is the face-area vector directed outward from cell $P$. This order of interpolation and normalisation specifies the conventional reference operator [22]; Eq. (8) provides its continuum notation. For uniform gravity, the modified pressure $p_{\rho gh} = p - \rho\, \mathbf{g} \cdot \mathbf{x}$ gives $-\nabla p + \rho \mathbf{g} = -\nabla p_{\rho gh} - (\mathbf{g} \cdot \mathbf{x}) \nabla \rho$ and yields the equivalent momentum balance:

$$\frac{\partial(\rho \mathbf{u})}{\partial t} + \nabla \cdot (\rho \mathbf{u} \otimes \mathbf{u}) - \nabla \cdot [2(\mu + \mu_{\mathrm{ML}}) \mathbf{S}^d] = -\nabla p_{\rho gh} - (\mathbf{g} \cdot \mathbf{x}) \nabla \rho + \sigma \kappa \nabla \alpha. \tag{10}$$

The three right-hand-side terms use face-normal gradients and the same cell-centre reconstruction operator, following the balanced-force principle for coupling pressure, gravity, and capillarity [27]. For zero-gravity capillary equilibrium, $\mathbf{u} = \mathbf{0}$ and the interior-to-exterior pressure jump is $\sigma/R$ for a circle or $2\sigma/R$ for a sphere of radius $R$. Section 3 replaces the curvature in Eq. (8) with a learned estimate while retaining every other component of the momentum balance in Eq. (10).

## 2.4 Interface-gated constitutive closure

The second learned component is a non-negative, interface-localised scalar viscosity that contributes the symmetric deviatoric stress $2\mu_{\mathrm{ML}} \mathbf{S}^d$ to the momentum balance. The network predicts a bounded dimensionless coefficient within a prescribed constitutive form whose algebraic structure enforces the properties established below independently of training. The local filter width is $\Delta = \sqrt{V/h_z}$ for an extruded two-dimensional cell of volume $V$ and depth $h_z$, and $\Delta = V^{1/3}$ in three dimensions. Using the clipped copy $\alpha_c = \min(1, \max(0, \alpha))$ exclusively within the closure, the network receives the following ordered vector of dimensionless local flow and material-property features:

$$\mathbf{q} = \left[\alpha_c, \ln\left(1 + \frac{\Delta^2 s}{\nu}\right), \ln\left(1 + \frac{\Delta^2 w}{\nu}\right), \Delta |\nabla \alpha|, \ln \frac{\rho_1}{\rho_2}, \ln \frac{\mu_1}{\mu_2}, \ln\left(1 + \frac{\sigma \Delta}{\rho \nu^2}\right)\right]^{\mathrm{T}}. \tag{11}$$

The entries comprise the phase fraction, logarithmically transformed strain and rotation Reynolds measures, grid-relative interface sharpness, logarithmic material-property contrasts, and a local Laplace-number measure. The training mean $m_i$ and positive scale $d_i > 0$ define each normalised feature $\hat{q}_i = (q_i - m_i)/d_i$ before evaluation by the single-hidden-layer network. The perceptron maps these normalised inputs to a bounded coefficient through hyperbolic-tangent hidden activations and a logistic output scaled by the prescribed upper bound $C_{\max} > 0$:

$$C = C_{\max}\, \varsigma(\mathbf{w}_2 \cdot \tanh(\mathbf{W}_1 \widehat{\mathbf{q}} + \mathbf{b}_1) + b_2),$$
$$\varsigma(z) = \frac{1}{1 + e^{-z}}. \tag{12}$$

The output is admitted only when $\mathbf{q}$ lies within the axis-aligned training envelope $\mathscr{E}_\varepsilon$, expanded by a relative padding margin $\varepsilon$. Accordingly, $C^\star = C$ inside the envelope and $C^\star = 0$ outside it and therefore provides a deterministic fallback without interpreting range membership as a calibrated uncertainty estimate [74–76]. The constitutive projection combines this guarded coefficient with an interface gate and the strain-based mixing-length scale $\Delta^2 s$, consistent with Smagorinsky-type eddy-viscosity scaling [54, 56]:

$$\begin{aligned} \nu_{\mathrm{ML}} &= \lambda \min(\chi\, C^{\star} \Delta^2 s,\, b_\nu\, \nu), \\ \chi &= 4\alpha_c(1-\alpha_c), \qquad \mu_{\mathrm{ML}} = \rho\, \nu_{\mathrm{ML}}. \end{aligned} \tag{13}$$

The strength parameter $\lambda\in[0,1]$ scales the correction, while $b_\nu=0.1$ limits the closure viscosity relative to the molecular mixture viscosity.

**Proposition 1. Bounds and locality.**

Since $0\le\chi\le1$ and $0\le C^{\star}\le C_{\max}$, the projection satisfies $0\le\nu_{\mathrm{ML}}\le\lambda\min(C_{\max}\Delta^2 s,\, b_\nu \nu)$ for every admissible state. The closure vanishes in either pure phase, in rigid motion with $s=0$, for $\lambda=0$, and outside the padded training envelope. For bounded strain, $\nu_{\mathrm{ML}}=\mathcal{O}(\Delta^2)$, so the correction vanishes under refinement and remains formally consistent with the uncorrected one-fluid equations. These statements follow directly from the bounded logistic output, non-negative projection, and interface gate and establish bounds and locality independently of the trained network weights.

**Proposition 2. Continuum dissipativity.**

The dimensionless features preserve Galilean invariance and invariance under fixed rotations and reflections, so the scalar viscosity retains these symmetries in the resulting stress. The following identity assumes zero closure-stress boundary work, as obtained at stationary no-slip walls and impermeable free-slip or symmetry boundaries with zero tangential closure traction. Paired periodic boundary contributions also cancel, so the closure-stress contribution to the kinetic-energy rate satisfies:

$$\begin{aligned} \int_\Omega \mathbf{u}\cdot\nabla\cdot\left(2\mu_{\mathrm{ML}}\mathbf{S}^d\right) \mathrm{d}V &= -\int_\Omega 2\,\mu_{\mathrm{ML}}\, \mathbf{S}^d{:}\mathbf{S}^d\, \mathrm{d}V \\ &\le 0. \end{aligned} \tag{14}$$

Integration by parts eliminates the boundary work under the stated conditions and yields the volume integral $-\int_\Omega 2\,\mu_{\mathrm{ML}}\mathbf{S}^d{:}\nabla\mathbf{u}\,\mathrm{d}V$. Symmetry and zero trace give $\mathbf{S}^d{:}\nabla\mathbf{u}=\mathbf{S}^d{:}\mathbf{S}^d\ge0$, so non-negative viscosity establishes the displayed bound independently of the trained network weights. This identity applies to the complete symmetric stress and therefore guides the discretisation of all gradient contributions beyond the scalar Laplacian $\nabla\cdot(\mu_{\mathrm{ML}}\nabla\mathbf{u})$. The identity establishes dissipativity of the continuum closure-stress contribution under the stated boundary conditions; discrete energy stability of the complete variable-density algorithm lies outside this proof. Energy-conserving closure architectures provide complementary single-phase guarantees [63, 64]. The scalar viscosity represents non-negative, strain-aligned dissipation and excludes backscatter or stress anisotropy independent of the resolved strain [58, 72]. The learned curvature component separately addresses capillary-force discretisation, while the momentum closure retains its bounded, non-negative constitutive structure throughout the solver comparison.

## 2.5 Training targets for the closure from filtered resolved fields

Training targets are extracted from resolved simulations by explicit filtering according to the Favre-filtered formulation of two-phase large-eddy simulation and its treatment of unresolved transport [81]. This construction relates the targets to coarse-grid variables and addresses consistency between *a priori* regression and *a posteriori* solver behaviour [62]. A non-overlapping top-hat filter over $f\times f$ fine cells defines the filtered density, Favre-filtered velocity, and convective subfilter stress through the following relations:

$$\bar{\rho}=\overline{\rho}, \qquad \tilde{\mathbf{u}}=\frac{\overline{\rho\mathbf{u}}}{\bar{\rho}}, \qquad \boldsymbol{\tau}=\overline{\rho\mathbf{u}\otimes\mathbf{u}}-\bar{\rho}\,\tilde{\mathbf{u}}\otimes\tilde{\mathbf{u}}. \tag{15}$$

Centred differences of $\tilde{\mathbf{u}}$ define the filtered deviatoric strain tensor $\tilde{\mathbf{S}}^d$ with the same trace convention as the constitutive model. Projecting $\boldsymbol{\tau}$ onto the modelled stress $-2\mu_{\mathrm{ML}}\tilde{\mathbf{S}}^d$ yields the scalar viscosity that minimises the local squared tensor error:

$$\mu^{*}=-\frac{\boldsymbol{\tau}:\tilde{\mathbf{S}}^d}{2\,\tilde{\mathbf{S}}^d:\tilde{\mathbf{S}}^d}. \tag{16}$$

The isotropic component of $\boldsymbol{\tau}$ does not contribute because the filtered strain tensor is traceless, so only the deviatoric stress enters the projection. Samples with negligible strain are discarded, while negative coefficients associated with local backscatter are counted and projected to zero to obtain a dissipative target. For retained coarse cells within the interface band and away from walls, the filtered-state gate is $\chi_{\mathrm{F}}=4\bar{\alpha}(1-\bar{\alpha})$ and the strain magnitude is $\tilde{s}=\sqrt{2\tilde{\mathbf{S}}^d:\tilde{\mathbf{S}}^d}$. Retained samples satisfy $0<\bar{\alpha}<1$, $\tilde{s}>0$, and $\bar{\rho}>0$, so normalisation by the local constitutive scale gives a well-defined bounded target:

$$y=\mathrm{clip}\left[\frac{\max(0,\mu^{*})}{\bar{\rho}\,\chi_{\mathrm{F}}\,\Delta^2\tilde{s}\,C_{\max}},\,0,\,1\right]. \tag{17}$$

The target specifies the nominal coefficient $C/C_{\max}$ independently of the deployment strength $\lambda$. Equation (13) subsequently applies the selected strength, feature-envelope guard, and viscosity cap to the inferred coefficient. Training minimises the regularised mean-squared error between $C/C_{\max}$ and $y$ through mini-batch Adam [92]. Normalisation constants and the feature envelope are computed exclusively from training samples, excluding validation and test trajectories from both feature scaling and guard construction. A hash of the physical parameters assigns whole source simulations to training, validation, and test partitions. This trajectory-level separation excludes snapshots and neighbouring cells from held-out simulations during training. Section 6.4 documents the training dataset for this convective-stress baseline, whose scope excludes the filtered compression, viscous, capillary, and pressure contributions to the coarse-grid residual [41, 67, 70, 71, 73].

# 3 Solver-consistent learned curvature

## 3.1 Curvature diagnosis and solver-consistent training data

The conventional capillary force is evaluated on mesh faces as $\sigma\,\bar{\kappa}_f\,\nabla_f^{\perp}\alpha$, where $\bar{\kappa}_f$ linearly interpolates the cell curvature defined in Eq. (8). We assess its contribution to static pressure balance through the following force-weighted curvature mean and associated signed pressure-jump estimate across the interface band:

$$\Delta p_h^{\mathrm{est}} := \sigma\,\langle\kappa_f\rangle_{\Delta\alpha}, \qquad \langle\kappa_f\rangle_{\Delta\alpha}=\frac{\sum_f \kappa_f\,|\Delta_f\alpha|}{\sum_f|\Delta_f\alpha|}. \tag{18}$$

Here, $\Delta_f\alpha$ denotes the volume-fraction difference across face $f$, and the sums include all faces that span the numerically resolved interface band. The sign follows increasing $\alpha$; for gas bubbles, the positive interior-to-exterior pressure-jump estimate is therefore $-\Delta p_h^{\mathrm{est}}=|\Delta p_h^{\mathrm{est}}|$. This global mean defines a diagnostic estimate whose agreement with the measured bulk pressure jump is assessed empirically. For constant face curvature and compatible static pressure balance between pure phases, it reproduces the signed Laplace jump exactly. Predicted magnitudes of 80.6, 82.5, and 83.4 compare with measured values of 81.5, 83.4, and 84.2 on the corresponding grids. Agreement within approximately 1.1 % supports force-weighted curvature bias as the leading contributor to the measured pressure deficit in these configurations. For solver-advected circular reference configurations, $\kappa/\kappa_{\mathrm{exact}}$ increases from approximately 0.35 on the gas side to 1.2 on the liquid side, as shown in Fig. 4(a). The divergence operator differentiates normalised gradients across the compressed two-to-three-cell profile and produces iso-contour curvature estimates within a strongly non-uniform discrete gradient field. Linear face interpolation favours lower magnitudes, so the force-weighted deficit remains 11–22 % over $R/h$=2.5–40 without observed improvement under refinement, as shown in Fig. 4(b).

This diagnosis motivates an estimator that assigns analytical reference curvature throughout the mixed-cell band and targets a capillary-force integral independent of numerical profile width. The solver generates the training stencils, so their phase-fraction distributions retain the effects of transport, compression, and parasitic currents encountered during deployment. Each gravity-free, doubly periodic square contains a circular gas bubble, a circular liquid droplet, or a flat gas slab inclined at a rational slope. Uniform initial velocities have random directions and speeds between 0.08 and 0.3, while 15 % of cases start at rest to sample stationary profiles. The continuum reference translates rigidly and supplies exact labels $\kappa=-1/R$ for bubbles, $\kappa=+1/R$ for droplets, and $\kappa=0$ for slabs under the convention of Eq. (8).

The dataset contains 68 circular cases that cover 17 radii with $R/h$=2.5–40, both benchmark fluid-property pairs, both phase arrangements, and three grid spacings. Eight additional slab cases provide zero-curvature examples at normal inclinations of 0°, 18.4°, 26.6°, and 45°. Training excludes the rising-bubble domain, gravitational forcing, and benchmark radius; the benchmark trajectories therefore provide independent evaluation despite the shared fluid-property pairs. Each case is integrated to $t$=0.5, and samples are extracted from every stored snapshot at cells that satisfy $10^{-3}<\alpha<1-10^{-3}$. Each sample comprises a cell-centred 5×5 volume-fraction stencil and the dimensionless label $h\kappa_{\mathrm{exact}}$ associated with the translating analytical reference.

### 3.2 Symmetry-preserving learning and closed-loop deployment

The network receives the normalised stencil $\mathbf{x}=2\boldsymbol{\alpha}-1\in\mathrm{R}^{25}$ in row-major order, so phase exchange corresponds to input sign reversal. For bounded phase fractions, its components lie in $[-1,1]$, while the symmetry construction remains valid for arbitrary real stencil values. A two-hidden-layer perceptron $g:\mathrm{R}^{25}\rightarrow\mathrm{R}$, with 128 hyperbolic-tangent units per layer, defines the dimensionless curvature prediction through antisymmetrisation and square-grid symmetry averaging:

$$h\kappa_{\mathrm{ML}}(\mathbf{x})=\frac{1}{8}\sum_{\mathbf{Q}\in D_4}\frac{1}{2}[g(\mathbf{Q}\mathbf{x})-g(-\mathbf{Q}\mathbf{x})]. \tag{19}$$

The group $D_4$ comprises the eight rotations and reflections of the square stencil, each of which permutes the input entries without altering their values. Antisymmetrisation makes the prediction exactly odd under phase exchange $\alpha\mapsto1-\alpha$, while group averaging enforces square-grid invariance for arbitrary network weights. These properties follow the symmetry-preserving approaches of Önder and Liu [85] and Cervone et al. [86], and the test suite verifies them to floating-point round-off. Training retains grid-scaled curvature as the response variable and uses mixed-cell weights consistent with subsequent band interpolation through the following relative squared-error objective:

$$\mathscr{L}=\frac{1}{N}\sum_n w_n\frac{[h\kappa_{\mathrm{ML}}(\mathbf{x}_n)-y_n]^2}{(|y_n|+\epsilon_n)^2},\qquad w_n=4\alpha_n(1-\alpha_n)+0.1. \tag{20}$$

The labels are $y_n=h\kappa_{\mathrm{exact}}$, with denominator floors $\epsilon_n$=0.03 for curved interfaces and $\epsilon_n$=0.3 for flat interfaces. The weights emphasise mixed cells, while the larger flat-interface floor limits the influence of zero-curvature samples, which constitute approximately one quarter of the dataset. This choice addresses the observed tendency of flat-interface samples to dominate the uniform-floor loss and suppress curved-interface predictions towards zero. Optimisation uses mini-batch Adam [92] with an initial learning rate of $2\times10^{-3}$ that is halved every 60 epochs; the lowest validation error determines the selected epoch. Each mini-batch receives one random $D_4$ transformation, and case-index grouping assigns whole simulations to training (60 %), validation (20 %), and test (20 %) partitions. All snapshots and neighbouring cells from held-out test trajectories remain outside training, so the partitions preserve independence at the simulation level.

During deployment, a precomputed Cartesian index map gathers the 5×5 stencil in each band cell through periodic wrapping on cyclic patches and zero-gradient clamping at walls. A dependency-free C++ kernel evaluates Eq. (19), after which division by $h$ converts the dimensionless prediction into the dimensional cell curvature. The solver replaces predictions with $|h\kappa_{\mathrm{ML}}|>0.5$ by the conventional estimate and records these events, which correspond to local curvature radii below two cells. For faces with $w_P+w_N>0$, the following band-weighted interpolation converts adjacent cell estimates into the face curvature used for capillary forcing:

$$\kappa_f=\frac{w_P\kappa_P+w_N\kappa_N}{w_P+w_N},\qquad w=\alpha(1-\alpha). \tag{21}$$

Here, $P$ and $N$ denote adjacent cells, and a face between mixed and pure phases receives the mixed-cell curvature without dilution by a bulk value. The same flux $\sigma\kappa_f\nabla_f^{\perp}\alpha|\mathbf{S}_f|$ enters the momentum predictor, Eq. (24), and pressure equation, Eq. (26), to preserve consistent pressure–capillary coupling [27]. Dictionary selection activates the learned estimator, while the conventional option retains the OpenFOAM-7 discretisation and reproduces its diagnostic histories to the precision reported in Section 6.1. For comparison, smoothed CSF evaluates Eq. (8) from an auxiliary phase fraction after $n_s$ passes of $\alpha\leftarrow\overline{\overline{\alpha}^f}$ and uses the same band-weighted face interpolation. Each pass averages the face-interpolated field back to cells according to standard CSF smoothing practice, with the smoothing width retained as a tunable parameter [26, 29, 31]. We use $n_s=8$ because it gives the best performance in the *a priori* comparisons and provides a tuned classical reference.

The deployed estimator also changes the distribution of phase-fraction profiles supplied to subsequent predictions, since its capillary force differs from the conventional training-data generator. This feedback motivates iterative data aggregation [90] and addresses the deployment-consistency concerns identified for learned constitutive closures [41, 67, 71]. The first iteration runs all cases with conventional curvature and trains $\mathcal{M}_1$ on the resulting solver-generated stencils and analytical reference labels. At iteration $k+1$, the same cases receive new random velocities and run with $\mathcal{M}_k$; the accumulated datasets then train $\mathcal{M}_{k+1}$. Each case retains its original training, validation, or test assignment across iterations, so related rollouts remain within the same partition of the aggregated dataset. A run is excluded as unstable if its maximum speed exceeds the imposed speed by more than 0.08 after $t=0.1$. The final estimator $\mathcal{M}_3$ uses 394 497 samples from 213 retained runs, and Fig. 1 summarises the complete training and deployment procedure.

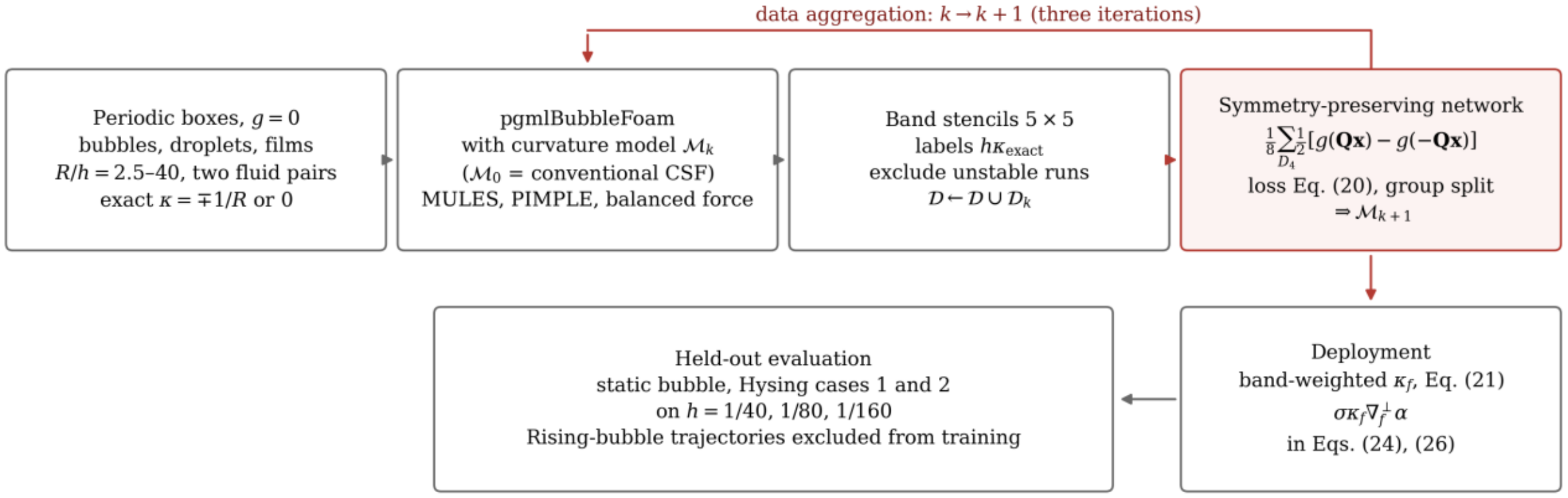

**Figure 1.** Solver-consistent curvature workflow: periodic reference configurations provide exact curvature labels for solver-generated volume-fraction stencils, and iterative data aggregation incorporates feedback between predictions and evolving profiles. The symmetry-preserving network enters the balanced-force coupling through band-weighted face interpolation; red marks learned components and update paths. The final model is evaluated on independent static-bubble and Hysing configurations, with rising-bubble trajectories excluded from training.

# 4 Numerical method and implementation

## 4.1 Finite-volume transport and pressure–velocity coupling

The governing equations use OpenFOAM's cell-centred finite-volume formulation on a collocated grid, with a common discretisation framework for phase transport, momentum, and pressure [21, 93, 94]. Time derivatives use first-order implicit Euler, momentum convection uses linear upwinding with Gauss-linear velocity gradients, and diffusion uses second-order Gauss-linear Laplacians. Phase-fraction transport combines van Leer-limited advective fluxes with linearly interpolated compressive fluxes, so interface advection and compression retain distinct numerical treatments. The multidimensional universal limiter for explicit solution (MULES) applies flux-corrected transport [17] through a bounded low-order flux and a limited antidiffusive correction:

$$\phi_{\alpha,f}=\phi_{\alpha,f}^{\mathrm{UD}}+\Lambda_f\left(\phi_{\alpha,f}^{\mathrm{HO}}-\phi_{\alpha,f}^{\mathrm{UD}}\right), \quad 0\leq\Lambda_f\leq 1. \tag{22}$$

The superscripts UD and HO denote upwind and higher-order phase fluxes, respectively, and iterative updates of the face limiters $\Lambda_f$ enforce local admissible bounds. Two phase-fraction correctors and two subcycles are used per time step, and the limited phase flux determines the mass flux supplied to momentum transport. The relation $(\rho\phi)_f=(\rho_1-\rho_2)\phi_{\alpha,f}+\rho_2\phi_f$ couples phase and mass transport consistently, which is particularly important at a density ratio of 1000 [23, 91].

Within this conservative transport framework, the closure stress follows the molecular-stress treatment in OpenFOAM-7 [22] and retains both symmetric-gradient contributions. An implicit Laplacian and an explicitly evaluated transpose term provide the following stress decomposition for the momentum update:

$$\begin{aligned} -\nabla\cdot\left(2\mu_{\mathrm{ML}}\mathbf{S}^d\right) &\approx -\nabla\cdot\left(\mu_{\mathrm{ML}}\nabla\mathbf{u}^{n+1}\right) \\ &\quad -\nabla\cdot\left[\mu_{\mathrm{ML}}\,\mathrm{dev}_2\left(\left(\nabla\mathbf{u}^{*}\right)^{\mathrm{T}}\right)\right], \\ \mathrm{dev}_2(\mathbf{A}) &= \mathbf{A}-\frac{2}{3}\mathrm{tr}(\mathbf{A})\mathbf{I}. \end{aligned} \tag{23}$$

The superscripts $n+1$ and $*$ denote the implicit velocity update and the lagged velocity used for the explicit stress contribution, respectively. Both gradient terms are retained to represent the complete symmetric stress assumed in Proposition 2, and the assembled momentum operator defines the predictor equation:

$$\mathscr{M}\mathbf{u}=\mathscr{R}\left[\left(\sigma\kappa_f\nabla_f^{\perp}\alpha-(\mathbf{g}\cdot\mathbf{x})_f\nabla_f^{\perp}\rho-\nabla_f^{\perp}p_{\rho gh}\right)\left|\mathbf{S}_f\right|\right]. \tag{24}$$

Here, $\mathscr{M}$ denotes the assembled momentum matrix, $\nabla_f^{\perp}$ is the face-normal gradient, and $\mathscr{R}$ reconstructs cell-centred vectors from face fluxes. Conventional curvature uses linear interpolation of Eq. (8), whereas the learned and smoothed models use the band-weighted interpolation of Eq. (21). Pressure–velocity coupling employs PIMPLE, which combines PISO [95] and SIMPLE with one outer iteration and three pressure correctors per time step. The decomposition $\mathscr{M}\mathbf{u}=a_P\mathbf{u}-\mathbf{H}(\mathbf{u})$ separates the diagonal coefficient $a_P$ from the remaining momentum contributions and defines the predicted and buoyancy–capillary fluxes:

$$\begin{aligned}
\phi_H&=\left(\frac{\mathbf{H}}{a_P}\right)_f\cdot\mathbf{S}_f+\phi_{\mathrm{ddt},f},\\
\phi_{\mathrm{ddt},f}&=I_f\left(\frac{\rho}{a_P}\right)\mathrm{ddtCorr}_f(\mathbf{u},\phi),\\
\phi_g&=\left(\frac{1}{a_P}\right)_f\left(\sigma\kappa_f\nabla_f^{\perp}\alpha-(\mathbf{g}\cdot\mathbf{x})_f\nabla_f^{\perp}\rho\right)\left|\mathbf{S}_f\right|.
\end{aligned} \tag{25}$$

For the present stationary, nonrotating meshes, $\mathrm{ddtCorr}_f$ denotes OpenFOAM's time-discretisation correction for consistency between cell velocities and face fluxes [22]. The interpolated inverse diagonal $(1/a_P)_f$ weights the buoyancy–capillary contribution, and the following modified-pressure equation enforces finite-volume continuity:

$$\nabla\cdot\left[\left(\frac{1}{a_P}\right)_f\nabla p_{\rho gh}\right]=\nabla\cdot\left(\phi_H+\phi_g\right). \tag{26}$$

The corrected flux $\phi=\phi_H+\phi_g-(1/a_P)_f\nabla_f^{\perp}p_{\rho gh}\left|\mathbf{S}_f\right|$ satisfies discrete continuity to solver tolerance. The identical capillary face quantity in the predictor and pressure equations preserves operator consistency, while stationary balance requires a compatible discrete pressure potential for the combined buoyancy–capillary force. A non-gradient force component can still drive parasitic circulation despite an accurate mean pressure jump; agreement in mean pressure therefore characterises only one aspect of discrete capillary balance. Equation (18) supplies the static pressure-jump diagnostic, while the reconstructed face contributions support assessment of local pressure–force balance. Learned curvature is updated after phase transport and mixture-property evaluation and remains fixed during pressure correction; no under-relaxation is applied within the coupling. The final pressure corrector uses diagonal-incomplete-Cholesky-preconditioned conjugate gradients with an absolute tolerance of $10^{-9}$, while the momentum and phase-fraction solvers specify symmetric Gauss–Seidel smoothing.

### 4.2 Time-step control and software implementation

To control temporal resolution consistently across curvature models, the adaptive time step satisfies convective and interface Courant limits of 0.25 and 0.2, respectively. The explicit capillary contribution also imposes the following stability bound, which supplements both Courant criteria and constrains the time step during weak-flow conditions:

$$\Delta t\leq\Delta t_\sigma=0.2\sqrt{\frac{(\rho_1+\rho_2)\,h^3}{\pi\sigma}}. \tag{27}$$

The $h^{3/2}$ dependence follows the capillary-wave time scale, and the prefactor provides a safety margin for explicit surface-tension forcing [26, 35]. A global cap of $\Delta t \le 2.5\times10^{-3}$ is additionally imposed, and the same Courant limits and capillary bound govern all three curvature models.

The implementation derives from the Foundation OpenFOAM-7 `interFoam` application [22]. Both networks use dependency-free, header-only C++ kernels in double precision; inference requires neither the external learning libraries nor coupling infrastructure used in other integration strategies [77–79]. Specifically, the kernels require no TensorFlow, PyTorch, ONNX, or embedded Python, while the dictionary `constant/curvatureProperties` selects the curvature model. If that dictionary is absent, conventional curvature remains active and existing cases retain their original solver configuration. Each model file begins with a magic string, network dimensions, and a provenance tag that identifies its expected format and declared training-data source. The tags `SOLVER_CONSISTENT_CURVATURE` and `FILTERED_CFD_STRESS` distinguish the curvature estimator from the viscosity closure and their respective data sources. Before inference, the parsers reject missing entries, non-finite values, trailing data, unsupported dimensions, and unknown provenance as part of the model-integrity checks.

The curvature kernel supports uniform single-layer Cartesian meshes in serial execution, detects periodic directions from cyclic patches, and rejects unsupported configurations with explicit diagnostics. At every time step, the solver records gas volume, centroid, mean bubble velocity, phase-fraction bounds, maximum velocity, closure viscosity, and curvature-fallback counts. The accompanying $L^1$ and $L^\infty$ norms of discrete divergence permit incompressibility errors to be assessed alongside the interface and momentum diagnostics. A standalone driver exercises the production inference headers, and all 33 unit and regression tests pass, with C++–NumPy curvature parity verified to relative tolerance $10^{-12}$. The suite also checks phase-exchange antisymmetry, $D_4$ invariance, corrupted-file rejection, and trained-model accuracy on exact circles across the tested configurations.

# 5 Geometry, configurations, and diagnostics

Following Hysing et al. [24], the benchmark domain is $[0,L]\times[0,2L]$ with $L$=1 m, as shown in Fig. 2(a). A single-cell extrusion of depth $h_z$=0.01 m uses `empty` conditions to suppress third-direction variation and represents the cross-section of a cylindrical bubble. The initially circular gas bubble has radius $R$=0.25 m and centre $(x_0,y_0)$=(0.5,0.5) m, with both phases initially at rest. Gravity acts in the negative $y$ direction with magnitude $g$=0.98 m s$^{-2}$, while horizontal and vertical walls impose no-slip and free-slip conditions, respectively. All walls impose zero normal gradients of $\alpha$, while the modified pressure uses fixed-flux conditions compatible with the prescribed wall velocities. The static configuration retains the first benchmark's domain, fluid properties, and initial bubble, with gravity set to zero to isolate capillary equilibrium.

These configurations use uniform meshes with $h/L$=1/40, 1/80, and 1/160, containing 3 200, 12 800, and 51 200 cells and resolving $R$ by 10, 20, and 40 cells, respectively. On each mesh, the initialisation utility classifies cells using their nearest and farthest corner distances, then integrates cut-cell gas fractions with 64×64 midpoint quadrature, as illustrated in Fig. 2(b). Relative initial-area errors remain below $10^{-3}$ on all grids, while the static modified pressure starts from the exact Laplace jump to isolate subsequent solver-induced deviations. The same utility initialises the periodic curvature-training domains of Fig. 2(c), which use cyclic boundaries in both directions and either benchmark fluid-property pair. Their side length is $8R$ for circular interfaces or an integer number of inclined-film periods, with the training configurations detailed in Section 3.1.

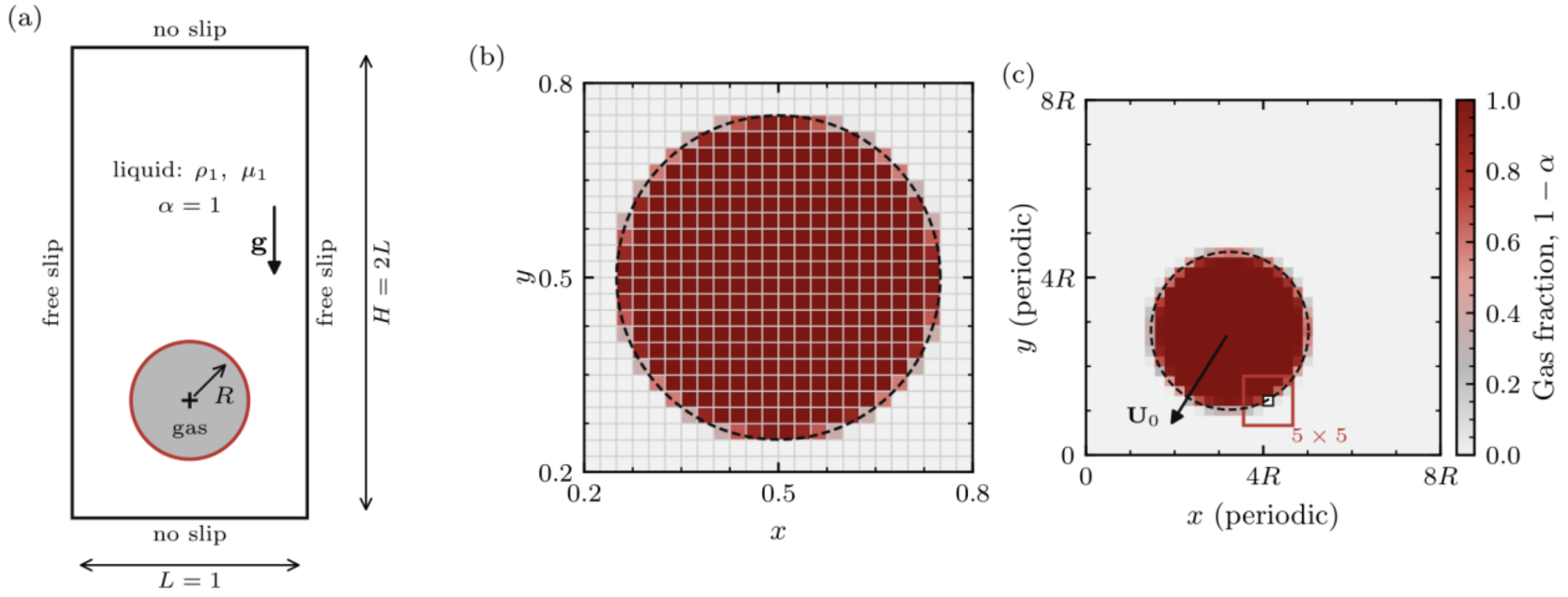

**Figure 2.** (a) The rising-bubble domain includes the phase labels and wall conditions, with a single `empty` cell of depth $h_z$ in the third direction. (b) The initial gas fraction $1{-}\alpha$ on the $h/L{=}1/40$ grid uses 64×64 subcell quadrature; dashed curves denote exact interfaces, and panels (b) and (c) share a colour scale. (c) The periodic, zero-gravity training box has side $8R$ and contains a bubble of radius $R{=}8h$ advected by uniform velocity $\mathbf{U}_0$. The snapshot at $t{=}0.5$ includes the rigidly translated reference interface and a 5×5 stencil centred on the band cell marked by an open square.

For the fixed rising-bubble geometry, the Reynolds and Eötvös numbers are $\mathrm{Re}{=}\rho_1\sqrt{gD}\,D/\mu_1$ and $\mathrm{Eo}{=}\rho_1 g D^2/\sigma$, respectively, based on the initial diameter $D{=}2R$ and liquid properties. Both cases have Re=35, while their Eötvös numbers are 10 and 125, respectively, as summarised with the physical parameters in Table 1. The density and viscosity ratios are both 10 in case 1 and increase to 1000 and 100, respectively, in case 2. The static case has $\mathrm{La}{=}\sigma\rho_1 D/\mu_1^2{=}122.5$; the exact pressure jump $\sigma/R{=}98\,\mathrm{Pa}$ provides an analytical reference for capillary equilibrium. Case 1 tests capillary control of compact-bubble deformation and drag, whereas case 2 combines weaker bulk capillary forces with dynamically important forcing in its thin skirts. The skirt curvature can approach the grid scale and the edge of the training range, and therefore tests the learned model near its resolution boundary.

**Table 1.** The table lists physical parameters and dimensionless groups for the three evaluation configurations, with dimensional quantities in SI units and gravity-based groups omitted for static equilibrium. The configurations are excluded from curvature-network training, although the same fluid-property pairs are represented in the periodic training cases described in Section 3.1.

| Quantity | Unit | Case 1 | Case 2 | Static |
|---|---|---|---|---|
| $\rho_1$, $\rho_2$ | kg m$^{-3}$ | 1000, 100 | 1000, 1 | 1000, 100 |
| $\mu_1$, $\mu_2$ | Pa s | 10, 1 | 10, 0.1 | 10, 1 |
| $\sigma$ | N m$^{-1}$ | 24.5 | 1.96 | 24.5 |
| $g$ | m s$^{-2}$ | 0.98 | 0.98 | 0 |
| Re, Eo | – | 35, 10 | 35, 125 | –, – |
| $\rho_1/\rho_2$, $\mu_1/\mu_2$ | – | 10, 10 | 1000, 100 | 10, 10 |
| End time | s | 3 | 3 | 1 |

The flow fields generated under these conditions are assessed using unmodified cell-centred values, without clipping $\alpha$ before the gas fraction $1{-}\alpha$ is evaluated. Bubble area, centroid, and mean rise velocity follow gas-volume weighting, with the extruded gas volume divided by $h_z$ to recover the cross-sectional area:

$$A=\frac{1}{h_z}\sum_k (1-\alpha_k)\, V_k,\quad y_c=\frac{\sum_k (1-\alpha_k)\, y_k V_k}{\sum_k (1-\alpha_k) V_k},\quad V_c=\frac{\sum_k (1-\alpha_k)\, u_{y,k} V_k}{\sum_k (1-\alpha_k) V_k}. \tag{28}$$

Here, $V_k$ is the cell volume, $y_k$ the cell-centre height, and $u_{y,k}$ the vertical velocity component, with sums taken over all computational cells. Alongside these volume-weighted quantities, circularity is $c=2\sqrt{\pi A}/P$, where $P$ sums the lengths of all closed $\alpha=1/2$ iso-contours and accounts for any detached gas components. The temporal evolution of these quantities is compared with $q_{\mathrm{ref}}$ using the following normalised $L^2$ measure over the rising-bubble interval $0\leq t\leq 3$:

$$e_{L^2}(q)=\left(\frac{\int_0^3 [q_h(t)-q_{\mathrm{ref}}(t)]^2\,\mathrm{d}t}{\int_0^3 q_{\mathrm{ref}}(t)^2\,\mathrm{d}t}\right)^{1/2}. \tag{29}$$

The time-history errors are evaluated at reference sample times, while scalar benchmark deviations are normalised by the tabulated TP2D values of Hysing et al. [24]. Reference histories and final shapes use the digitised datasets of Hysing et al. [24] and Zahedi, Kronbichler, and Kreiss [6] distributed with Lethe [96]. For the complementary static test, the pressure jump subtracts the mean pressure in $r>3R/2$ from that in $r<R/2$, with $r$ measured from the bubble centre. Parasitic motion is quantified by $\mathrm{Ca}_{\max}=\mu_1|\mathbf{u}|_{\max}/\sigma$, which expresses residual motion relative to surface tension through the liquid viscosity and domain-maximum speed. Interface diagnostics report the force-weighted face mean of Eq. (18), normalised by the exact curvature, together with its relative standard deviation across the interface band.

# 6 Results and discussion

## 6.1 Solver verification and curvature assessment

Baseline verification compares stock OpenFOAM-7 `interFoam` with PGMLBUBBLEFOAM for case 1 at $h=1/80$, with conventional curvature and the momentum closure disabled. All 13 diagnostics agree to the last printed digit at 301 output times over $0\leq t\leq 3$, which verifies the reference path without a curvature dictionary. With conservative phase transport retained, area drift remains below $3\times10^{-8}$, with case-1 maxima of $4.2\times10^{-9}$, $6.2\times10^{-9}$, and $2.9\times10^{-8}$ in Fig. 3. Case-2 maxima are $2.8\times10^{-9}$, $1.1\times10^{-8}$, and $2.0\times10^{-8}$; sign-changing residuals remain comparable with the conventional solver, while $\alpha$ stays within $[0,1]$ to $10^{-7}$ at every time step. The guard $|h\kappa_{\mathrm{ML}}|>0.5$ never activates in the case-1 core $0.05<\alpha<0.95$; case-2 activation remains confined to the formed skirts below the centroid. These lateral cells carry 8.8, 4.3, and 2.6 % of the gradient-weighted interface on successive grids, so their contribution decreases as the filaments become better resolved.

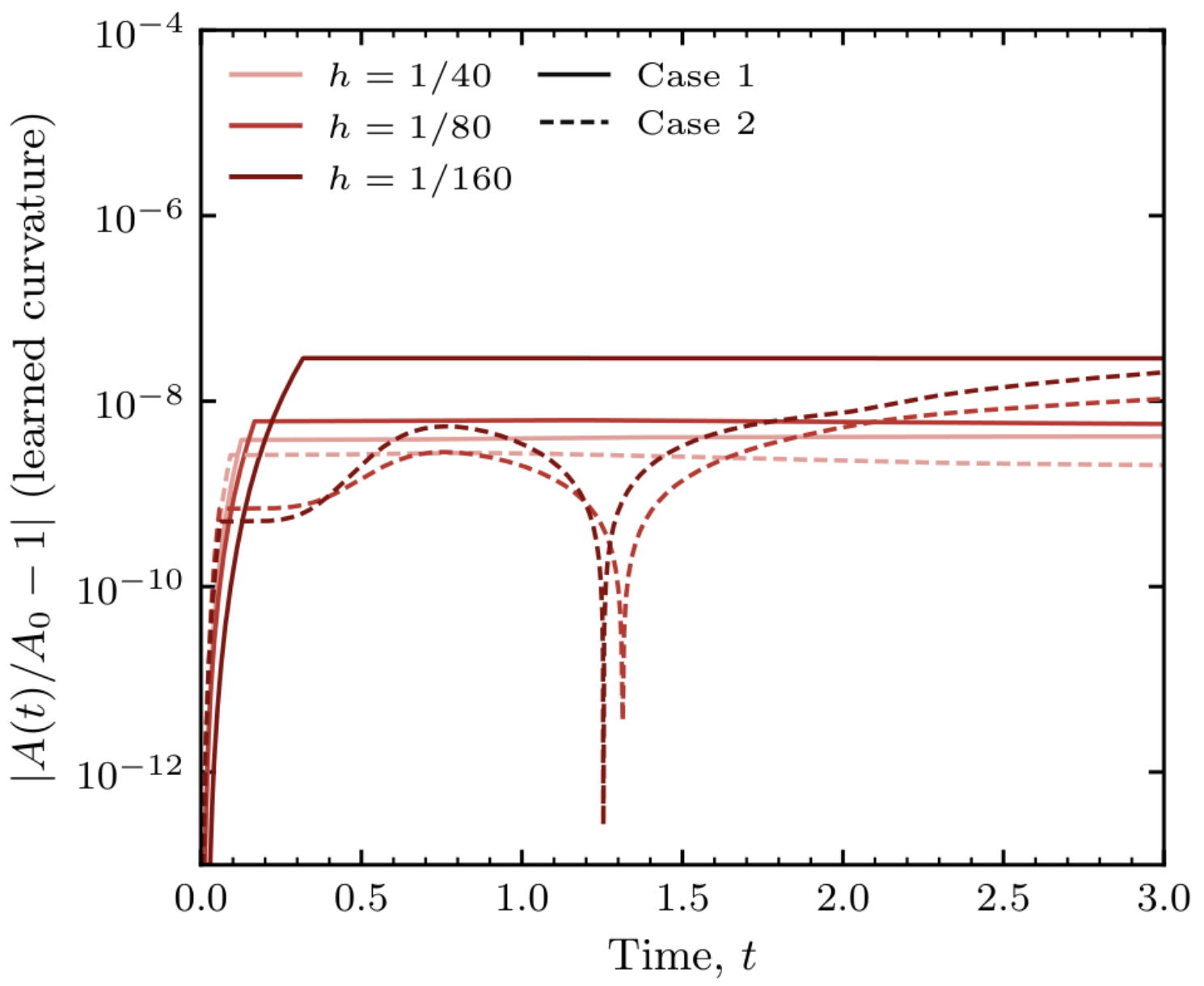

**Figure 3.** Relative bubble-area drift $|A(t)/A_0-1|$ is shown for all learned-curvature runs, with solid lines denoting case 1 and dashed lines case 2. Colour darkens with refinement, and cusps identify zero crossings of the signed drift before the absolute value is applied for display.

Against this verified baseline, curvature is assessed on 15 held-out bubble and droplet simulations spanning both fluid-property pairs and radii of 2.5–40 cells, excluded from training and model selection. Conventional cell estimates in Fig. 4(a) reach 34–48 % of exact curvature on the gas side and 56–68 % nearer the half level. After crossing the exact value near $\alpha$=1/2, they overshoot by more than 20 % and return to approximately 68 % in the outermost liquid cells. Learned bin means remain 0.94–0.99 and approach the target of Section 3.1, with interquartile widths of a few percent compared with broad conventional scatter. Face statistics in Fig. 4(b) use final conventional-solver snapshots, with linear interpolation for conventional curvature and band weighting for the learned and smoothed alternatives. Conventional deficits increase from 11.4 % at $R$=2.5$h$ to 18–22 % above six cells, with relative scatter rising from 0.3 to 1.9 at $R$=40$h$. Learned mean ratios of 0.970–1.002 and relative standard deviations of 0.06–0.26 compare with 0.969–1.009 and 0.05–0.12 for eight-pass smoothing. The approximately 2 % large-radius learned deficit accompanies an 8 % held-out relative root-mean-square error. These results motivate deployment assessment alongside mean accuracy, with trained weights and smoothing settings fixed throughout the comparisons.

Despite similar a priori accuracy, first-generation deployment exposes model-induced profile changes in the 76 periodic cases evaluated over 0.1≤$t$≤0.5 in Fig. 4(c). Conventional median and maximum excess speeds are $9.6\times10^{-3}$ and $4.9\times10^{-2}$, while the first learned model reduces the median to $5.8\times10^{-3}$. Nevertheless, 15 runs exceed 0.08 and reach 0.33 with sustained growth, consistent with the feedback-induced distribution shift described in Section 3.2. One aggregation step lowers the median to $3.6\times10^{-3}$, the 90th percentile from 0.15 to $1.1\times10^{-2}$, and the maximum to $7.1\times10^{-2}$. All runs then remain below the threshold, and only one exceeds 0.03, compared with 16 conventional runs at the same level. The threshold screens sampled trajectories; the final third-generation model pools three iterations with the 15 unstable runs excluded and case partitions fixed.

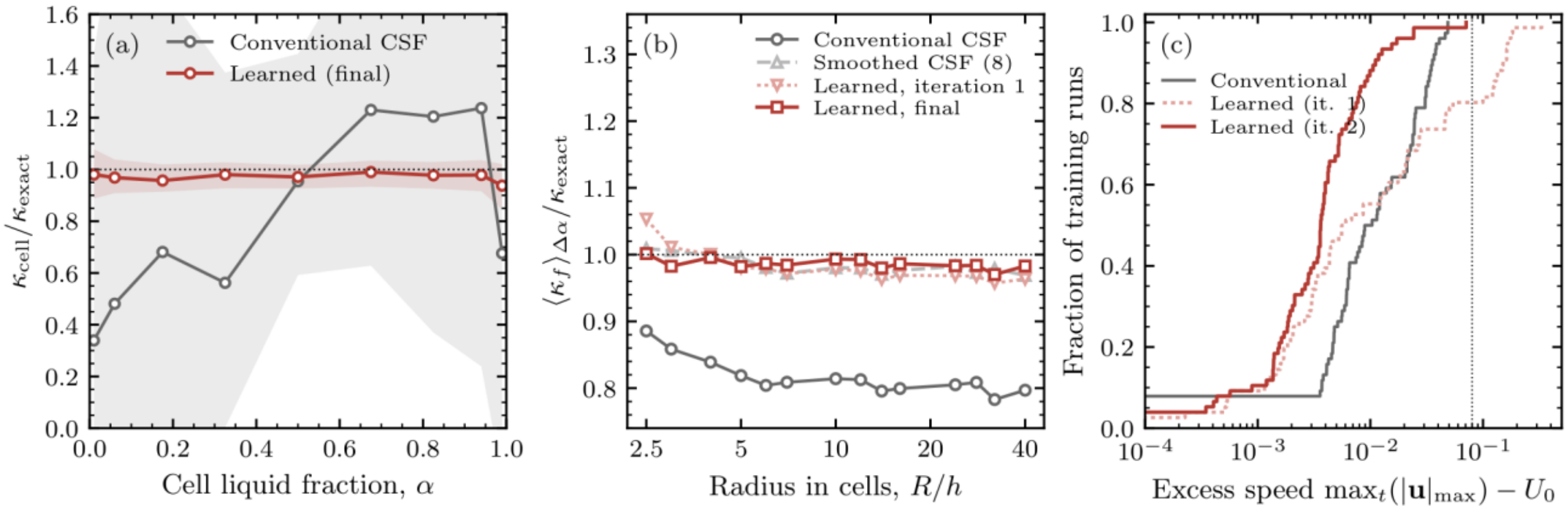

**Figure 4.** Curvature assessment uses held-out solver-generated translating circles, with panel (a) showing normalised cell curvature against liquid fraction; lines are bin means and shading spans the interquartile range. Panel (b) gives the normalised force-weighted face mean against radius in cells for conventional CSF, eight-pass smoothing, and first- and final-generation learned curvature. Panel (c) compares cumulative excess-speed distributions across the 76 periodic cases for conventional curvature and the first two learned generations, with the dotted line marking 0.08.

## 6.2 Static capillary equilibrium and force coherence

With transport conservation established, the static bubble isolates capillary equilibrium against the analytical pressure jump $\sigma/R$=98 and the zero-velocity reference state. Figure 5(a) compares horizontal-diameter pressure profiles at $t$=1 on the $h$=1/80 grid, with transitions confined to two or three cells. The conventional plateau is 83.4, or 14.9 % below the reference, whereas learned and smoothed curvature give 95.6 and 95.9, with errors of 2.4 and 2.1 %. Starting from the exact pressure jump, the conventional solution approaches its lower plateau within a few hundredths of a time unit, while the corrected models remain within approximately 5 %. Across the three grids, conventional pressure-jump errors are 16.9, 14.9, and 14.1 %, consistent with the persistent deficit identified by the force-weighted diagnosis of Section 3.1. Learned errors are 5.9, 2.4, and 5.5 % and yield approximately three- to sixfold reductions, although this configuration exhibits no monotonic refinement trend. The coarse- and fine-grid deficits exceed the 1–3 % indicated by the corresponding a priori mean-curvature assessment and reveal a difference between offline accuracy and stationary deployment. The stationary evaluation extends beyond the training horizon of $t$=0.5, and 15 % of the training cases start at rest. Eight-pass smoothing gives the smallest static pressure-jump errors, 1.2, 2.1, and 2.2 %, across the three grids.

The parasitic-current histories in Fig. 5(b) provide a complementary measure of local force balance beyond the mean pressure jump. At $h$=1/80, conventional $\mathrm{Ca}_{\max}$ rises to approximately $1.1\times10^{-2}$ and slowly approaches $6\times10^{-3}$ over the reported interval. Its averages over $0.5\le t\le 1$ are $1.1\times10^{-2}$, $6.3\times10^{-3}$, and $5.6\times10^{-3}$ on the three grids. Learned averages are $6.9\times10^{-4}$, $2.2\times10^{-3}$, and $2.0\times10^{-3}$, corresponding to reported reduction factors of approximately 16, 2.8, and 2.9. Smoothed averages are $1.3\times10^{-3}$, $8.3\times10^{-4}$, and $6.9\times10^{-4}$. Learned currents are therefore smallest on the coarse grid but approximately two to three times larger than smoothed currents on the finer grids. An extended $h$=1/80 calculation characterises the subsequent evolution after the slight current increase between $t$=0.4 and $t$=1. The capillary number peaks at $2.9\times10^{-3}$ near $t$=1.3, then falls to $1.9\times10^{-3}$ at $t$=2 and $1.2\times10^{-3}$ at $t$=3. The runtime pressure jump remains between 92.6 and 96.3, and the subsequent current decay supports bounded relaxation over the tested interval. Together, the pressure and current trends in Fig. 5(c) quantify improvements in mean capillary balance and document residual circulation during learned-curvature deployment.

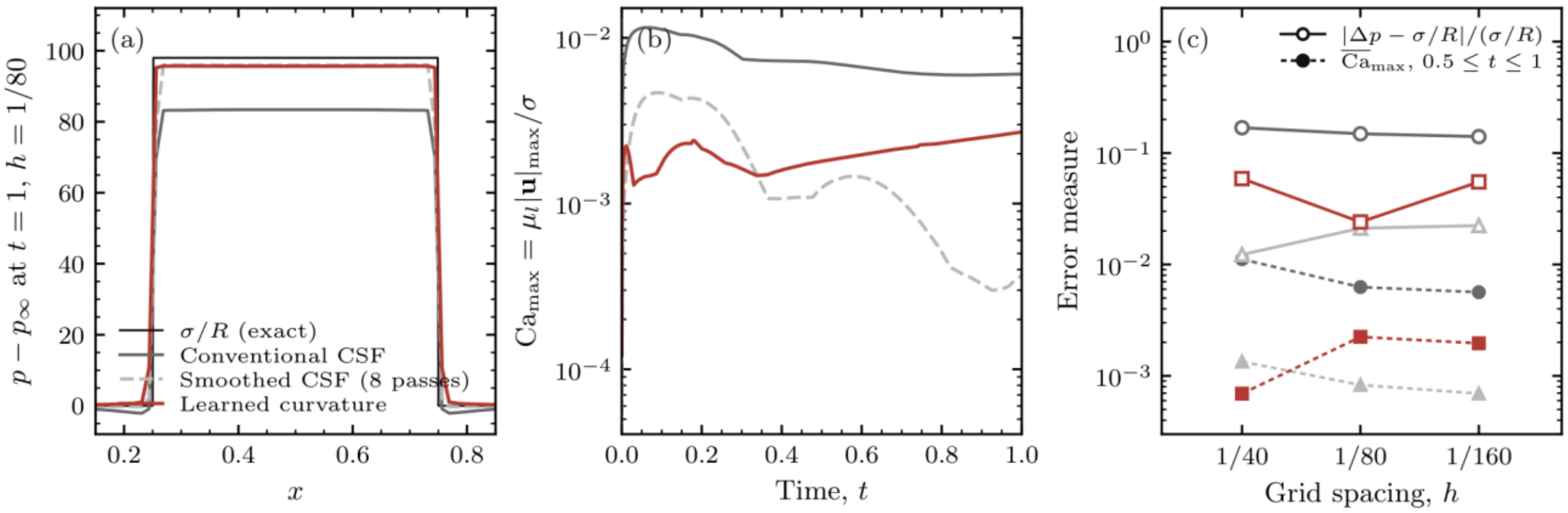

**Figure 5.** The zero-gravity static bubble has exact pressure jump $\sigma/R$=98, and the three models are compared through pressure profiles, current histories, and grid dependence. Panel (a) shows pressure relative to the far-field liquid along the horizontal diameter at $t$=1 and $h$=1/80, including the exact Laplace step. Panel (b) gives $\mathrm{Ca}_{\max}=\mu_1|\mathbf{u}|_{\max}/\sigma$ on the same grid; panel (c) compares pressure-jump errors at $t$=1 with currents averaged over $0.5\leq t\leq 1$. Open symbols and solid lines denote pressure errors, while filled symbols and dashed lines denote averaged currents plotted against grid spacing. Circles, triangles, and squares identify conventional, smoothed, and learned curvature, respectively, and provide the model encoding for the grid comparison in panel (c).

The spatial fields in Fig. 6 clarify the different current amplitudes by separating large-scale recirculation from motion concentrated immediately around the interface. Conventional curvature produces 16 recirculation cells with Cartesian fourfold symmetry [28–30], with capillary numbers above $10^{-3}$ extending more than one radius beyond the interface. Smoothing retains the symmetric pattern at almost an order-of-magnitude lower amplitude, while learned curvature largely suppresses the outer recirculation and gives a similarly quiet far field. The learned solution retains a thin interfacial current layer and cell-scale modulation near grid-aligned segments, where neighbouring stencils sample the discrete profile differently. The interfacial modulation is consistent with the larger a priori face-curvature scatter and the residual currents observed during static deployment.

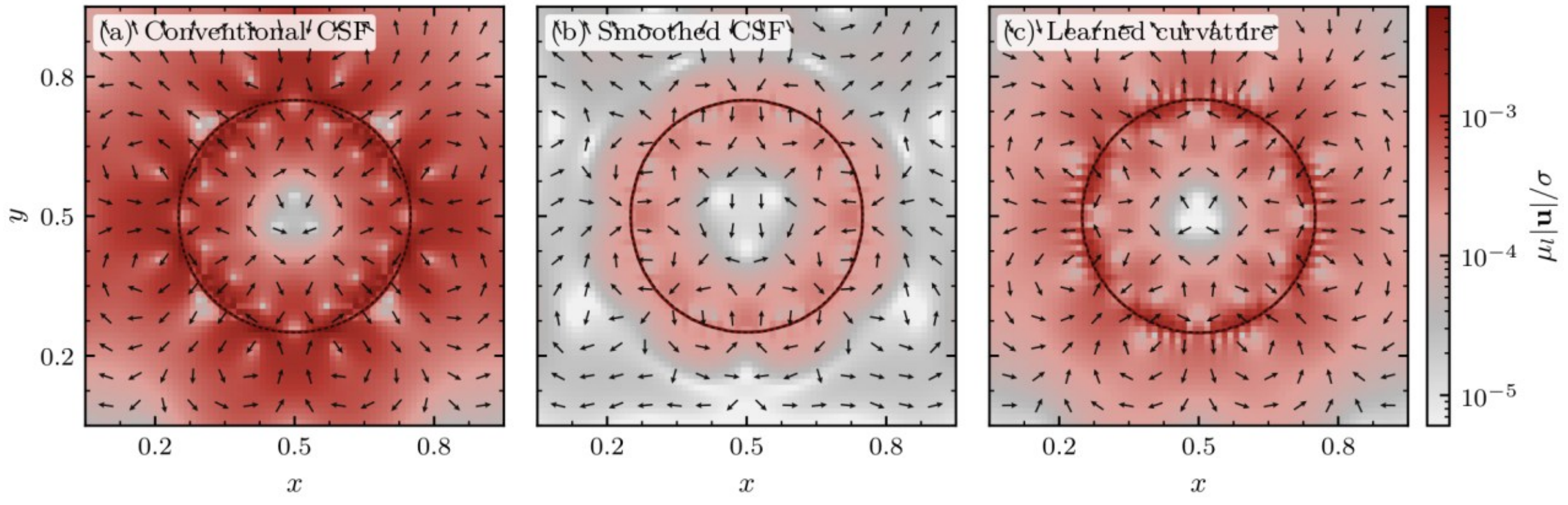

**Figure 6.** Static-bubble currents at $t$=1 and $h$=1/80 are shown for (a) conventional, (b) smoothed, and (c) learned curvature on a common logarithmic capillary-number scale. Colour represents $\mu_1|\mathbf{u}|/\sigma$, while uniform-length arrows show flow direction; the dark red contour is $\alpha$=1/2 and the dashed circle is the exact interface.

### 6.3 Rising-bubble dynamics and grid dependence

The rising-bubble tests extend the static capillary assessment to buoyancy-driven deformation and motion, with the compact bubble of case 1 shown in Fig. 7. Learned final centroid heights are 1.0776, 1.0776, and 1.0785, compared with 1.0703 for the finest conventional run and the TP2D value 1.0813. Conventional peak velocities of 0.2368, 0.2369, and 0.2364 retain deficits of approximately 2.0–2.2 % across the three grids. Learned peak velocities increase to 0.2398, 0.2402, and 0.2405, with reported deficits of 0.79, 0.64, and 0.51 %. Their peak times, 0.94, 0.93, and 0.92, approach the reference 0.9213 with refinement at the simulation output resolution of 0.01.

After the maximum, learned velocities follow the reference terminal behaviour, while the conventional terminal velocity remains approximately 3 % below the reference markers. Learned circularity minima of 0.8986, 0.8982, and 0.8986 lie within 0.35 % of the TP2D value 0.9013, compared with conventional deviations of 0.5–0.9 %. The subsequent circularity recovery also follows the reference more closely, which links the improved rise history to a more accurately maintained compact shape. Smoothing gives similar peak-velocity deficits of 0.60, 0.57, and 0.56 %, but its final centroid decreases from 1.0774 to 1.0750 and 1.0744. Its time-integrated velocity error increases from 0.80 to 1.19 and 1.38 %, whereas learned errors decrease from 1.04 to 0.82 and 0.65 %.

These observations support a capillary contribution to the conventional velocity bias: at Eo=10, curvature underestimation of 15–20 % is consistent with additional flattening and slower rise. The comparison retains coupled effects from phase transport and face interpolation, which also influence the resulting bubble shape and motion. The smoothing trend is consistent with a fixed-width kernel in cell units changing its mean-curvature bias as $R/h$ increases, while learned case-1 histories improve across the tested grids.

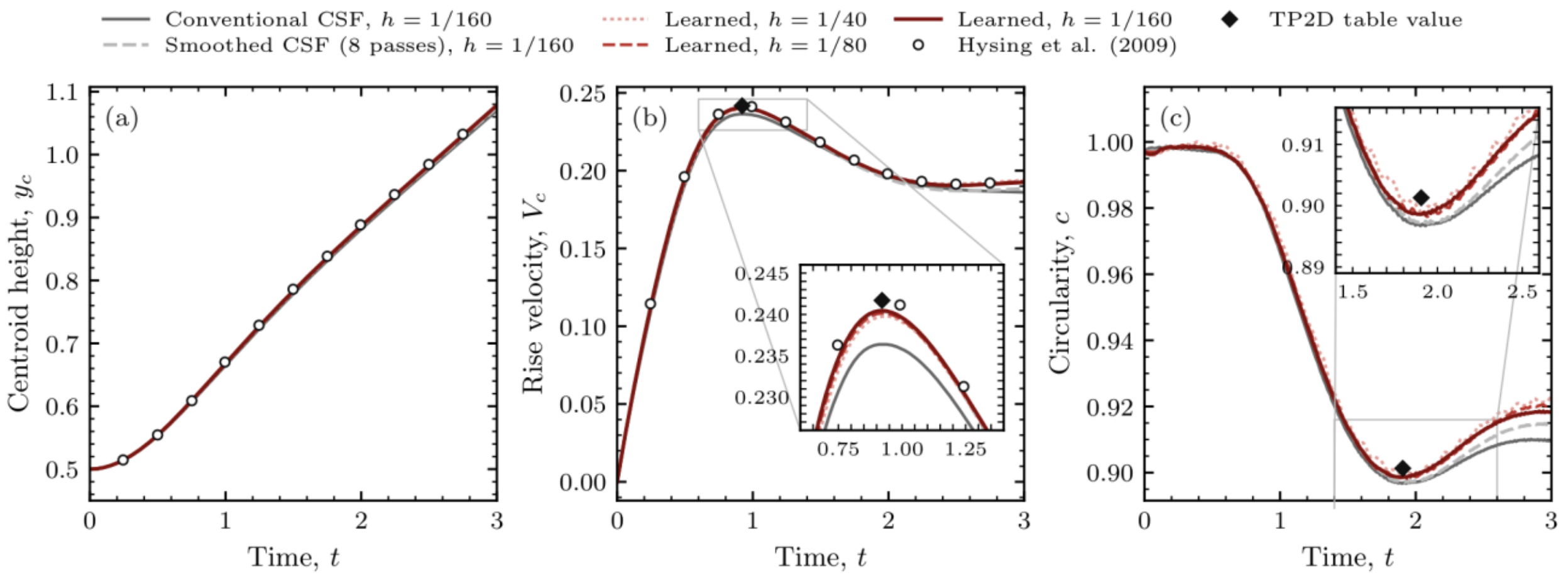

**Figure 7.** Case 1 (Re=35, Eo=10) compares (a) centroid height, (b) rise velocity, and (c) circularity for the three curvature models. Learned results use three grids, whereas conventional and smoothed results use $h$=1/160; open circles denote digitised Hysing et al. [24] references and diamonds TP2D scalars. The insets resolve the velocity maximum and circularity minimum, which complement the full histories used to assess transient and late-time agreement.

Case 2 extends the comparison to a density ratio of 1000 and weaker surface tension. The resulting skirted interfaces and double velocity maximum are shown in Fig. 8. Learned final centroids span 1.1112–1.1206, compared with conventional 1.1034–1.1147 and the TP2D value 1.1380, with improvement on every grid. The learned first maxima are 0.2501, 0.2493, and 0.2477 at $t$=0.74, or approximately 0.9–1.9 % below the reference 0.2524. Conventional first-peak deficits are 1.5–2.2 %, while learned second maxima increase from 0.2239 to 0.2314, approaching the reference 0.2434 more closely than conventional 0.2210–0.2298. The largest remaining velocity discrepancy occurs after $t$≈2, although learned terminal velocities remain closer to the reference on all three meshes.

The case-2 time-integrated velocity errors decrease from conventional 8.72, 7.02, and 5.64 % to learned 7.61, 5.75, and 4.72 %. Centroid errors similarly decrease from 1.66, 1.36, and 1.20 % to 1.23, 0.93, and 0.84 %, so both integrated diagnostics improve on every grid. Smoothing gives velocity errors of 9.35, 6.91, and 5.50 %. The coarse-grid result deteriorates, while the finer results improve by only 1.6 and 2.5 % in relative terms. The learned correction therefore provides the more consistent case-2 dynamical improvement, while substantial late-time discrepancies remain associated with the developing skirts.

Circularity falls to approximately 0.48–0.50 at $t$=3 for all models, but the reference groups disagree after breakup [24]; the TP2D minimum 0.5869 is therefore excluded from quantitative assessment. Learned runs develop up to 3, 5, and 4 disconnected gas regions, whereas conventional runs retain one connected bubble on the coarser grids and detach satellites only on the finest. The skirt-localised guard activity reported in Section 6.1 identifies where variable curvature approaches the interface-band scale and the constant-curvature training representation becomes least direct. The late-time comparison therefore reflects the combined learned estimator and the documented conventional fallback in the skirts.

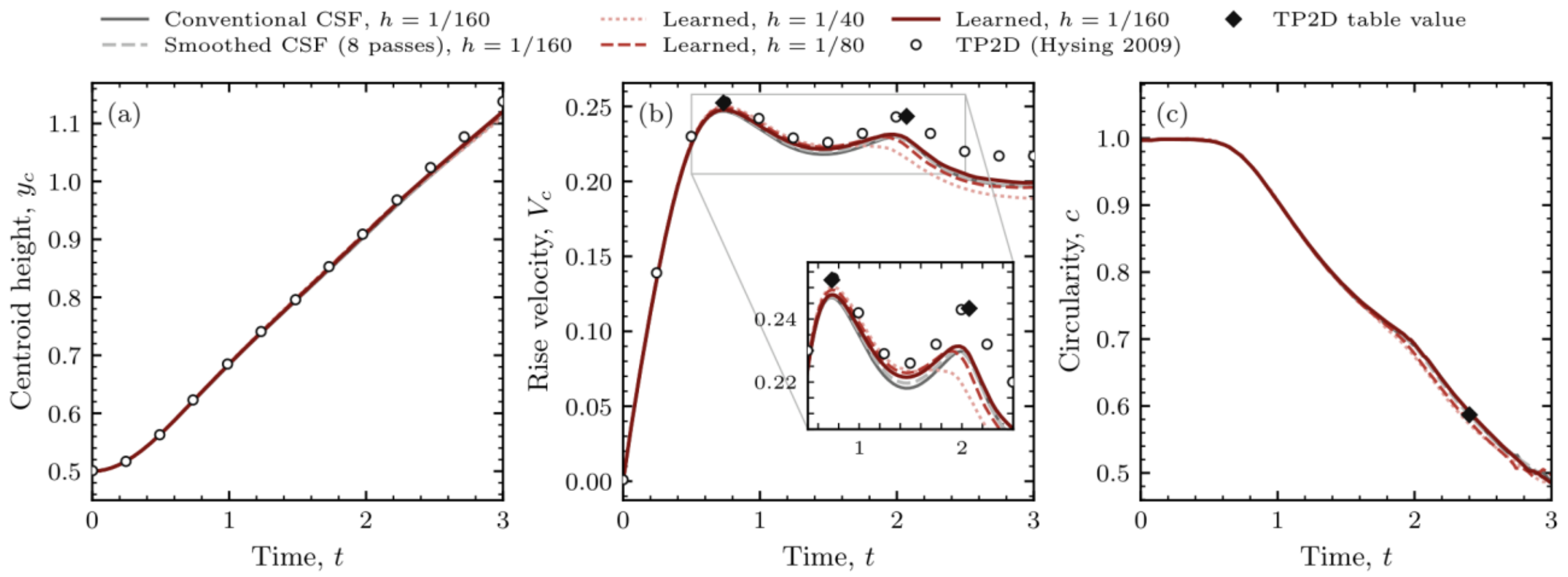

**Figure 8.** Case 2 (Re=35, Eo=125, $\rho_1/\rho_2$=1000) compares (a) centroid height, (b) rise velocity, and (c) circularity, with an inset resolving both velocity maxima. Learned results use three grids, while conventional and smoothed results use $h$=1/160; open circles denote digitised TP2D histories and diamonds the tabulated benchmark scalars.

The interface histories and vorticity fields in Fig. 9 connect these temporal diagnostics to the surrounding flow at the finest resolution. In case 1, the bubble forms an ellipsoidal cap with opposite-signed shear layers along the upper interface; these layers separate near the rim and diffuse into the wake. Case 2 develops a dome and nearly flat base joined by thin, almost vertical skirts, with satellite bubbles appearing near $y$≈0.6. Vorticity extends along the inner skirts and is nearly twice that of case 1, consistent with weaker capillary resistance and stronger local velocity gradients.

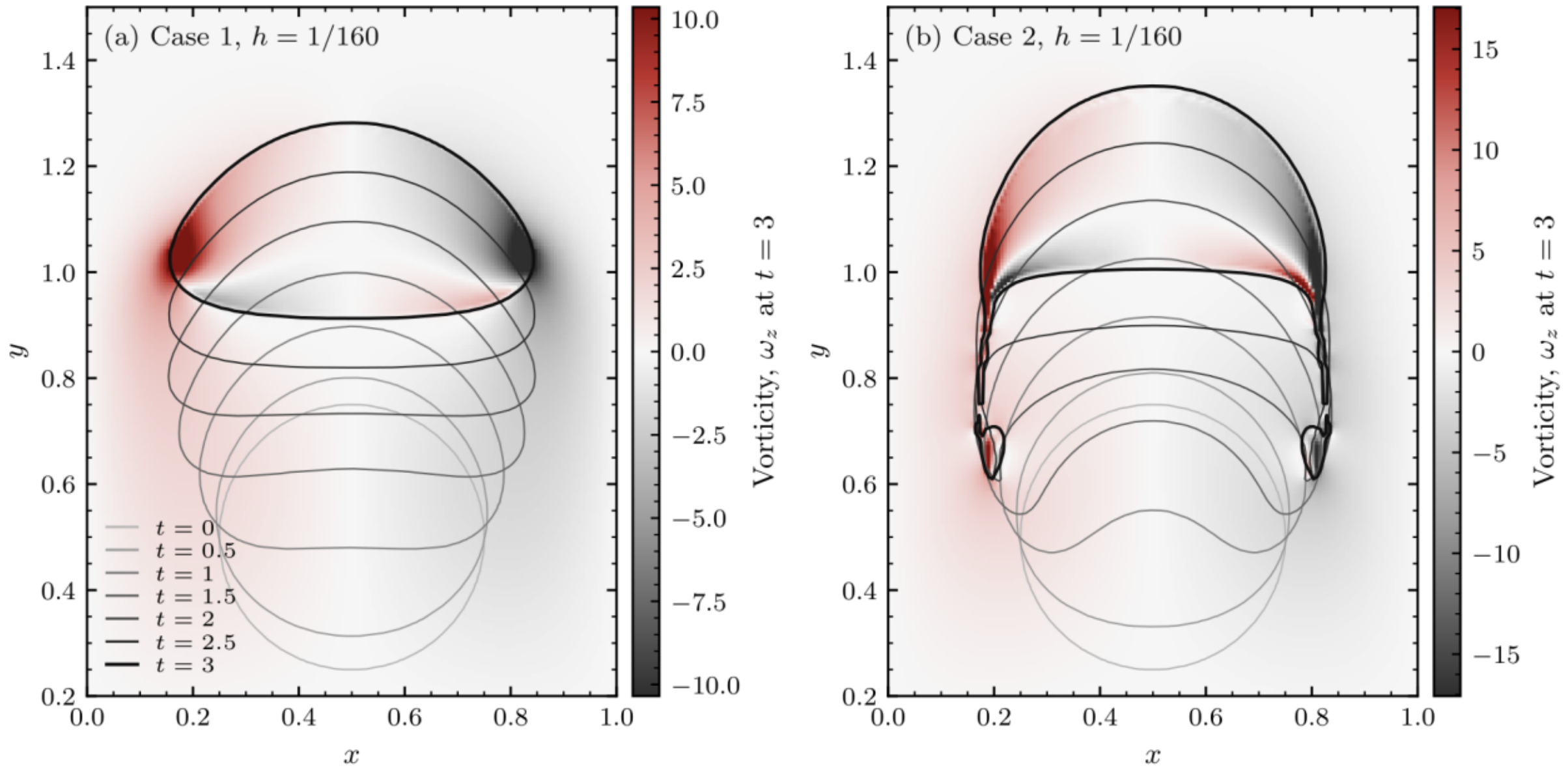

**Figure 9.** Interface contours at intervals of 0.5 time units are superimposed on $\omega_z$ at $t$=3 for learned curvature at $h$=1/160, with (a) case 1 and (b) case 2. Contours darken from light grey to the black final interface; the zero-centred colour map denotes counter-clockwise rotation in red and clockwise rotation in grey.

The final contours in Fig. 10 provide a complementary comparison of interface location against the published level-set and finite-element benchmark solutions. For case 1, the finest conventional contour lies up to approximately 0.015 below the Zahedi et al. [6] reference along the top and lower flanks. The smoothed contour lies between these curves, while learned contours agree with the reference within the plotted line width around the perimeter. In case 2, the computed domes closely follow the TP2D and MooNMD contours [24]; the learned base is slightly high at $h$=1/80 and matches at $h$=1/160. Learned satellite locations are consistent with the detached reference components, while grid-scale skirt thickness makes their detailed topology sensitive to numerical resolution [7, 24, 25].

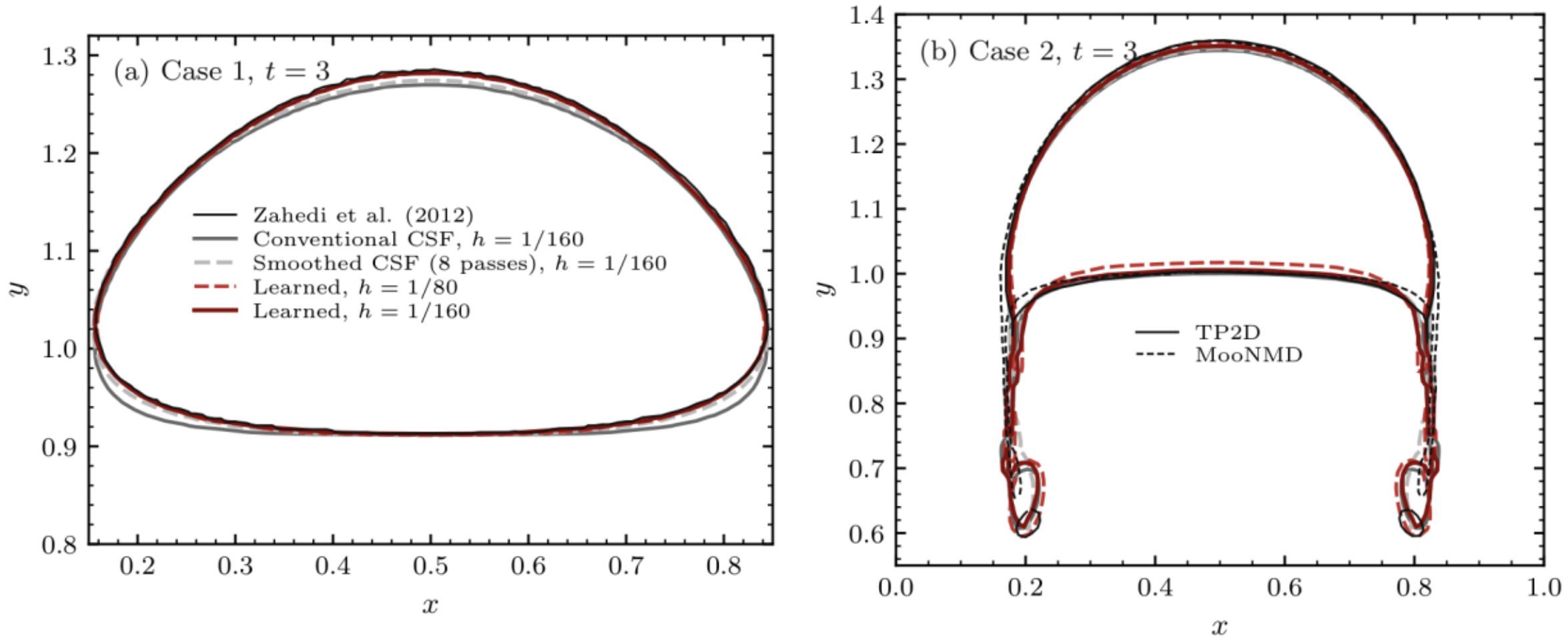

**Figure 10.** Interfaces at $t$=3 are compared with digitised references: (a) case 1 against Zahedi et al. [6], and (b) case 2 against TP2D and MooNMD [24], with detached gas components included. Conventional and smoothed models use the finest grid, while learned contours use the two finer grids; panel (b) retains the line styles of panel (a).

The bubble-frame sequence in Fig. 11 further illustrates skirt formation through the liquid velocity relative to the centroid at four instants. At $t$=0.75, the flow resembles that around a bluff body, with two symmetric internal circulation cells and a stagnation region below the bubble. By $t$=1.5, the stream passing the equator turns around the rear and raises the base into a crescent with downward-entrained tips. At $t$=2.25 and $t$=3, these tips form thin skirts enclosing a liquid recirculation zone, while internal gas circulation persists in the dome. The sequence is consistent with external shear opposed by capillary retraction and explains the sensitivity of skirt length and detachment to local curvature treatment.

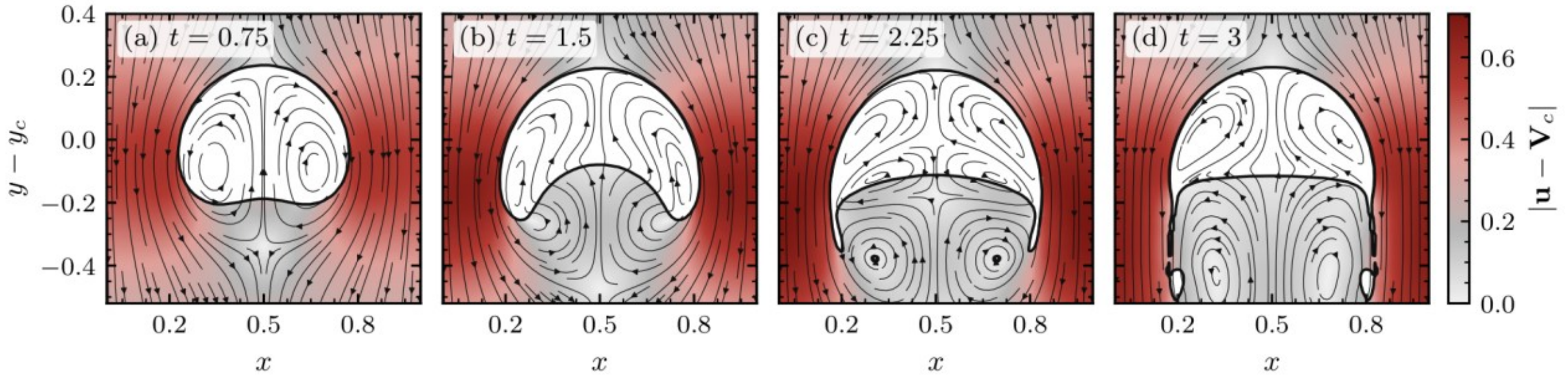

**Figure 11.** Case 2 is shown in the centroid frame with learned curvature at $h$=1/160 for (a) $t$=0.75, (b) $t$=1.5, (c) $t$=2.25, and (d) $t$=3. Colour gives liquid speed relative to $\mathbf{V}_c$, with the gas region white; black streamlines represent relative velocity and the thick contour denotes $\alpha$=1/2. The vertical coordinate is measured from the instantaneous centroid, so skirt development and the attached recirculation can be compared in the moving frame.

Figure 12 and Table 2 consolidate the grid trends, with conventional time-history errors decreasing at observed orders of approximately 0.15–0.32. Its first-peak deficit remains near 2 % in both cases, consistent with a persistent capillary contribution over the tested resolution range. For case 1, learned centroid and velocity errors improve by factors of 2.2–3.2, and the peak-velocity deficit falls below approximately 0.8 %. For case 2, relative improvements are 13–18 % for velocity and 26–32 % for centroid, with learned velocity-error orders of 0.40 and 0.28. These are comparable with conventional orders of 0.31 and 0.32, so the correction reduces error magnitude without establishing a higher asymptotic convergence order.

The first case-2 velocity maximum moves away from the reference with refinement for every model, consistent with compensating discretisation effects on coarser grids. Smoothed case-2 history errors decrease with refinement but remain close to conventional results on finer grids, whereas its case-1 velocity error increases. Across the 18 reported combinations of case, grid, and metric, learned curvature gives the smallest error in 14, smoothing in 4, and conventional curvature in none. The smoothing wins comprise all three case-1 metrics on the coarsest mesh and its first-peak deviation on the intermediate mesh, which demonstrates the configuration dependence of the comparison. The finest mesh remains one refinement level coarser than the original benchmark's finest grid, and history references are digitised, while TP2D scalar values retain their tabulated provenance.

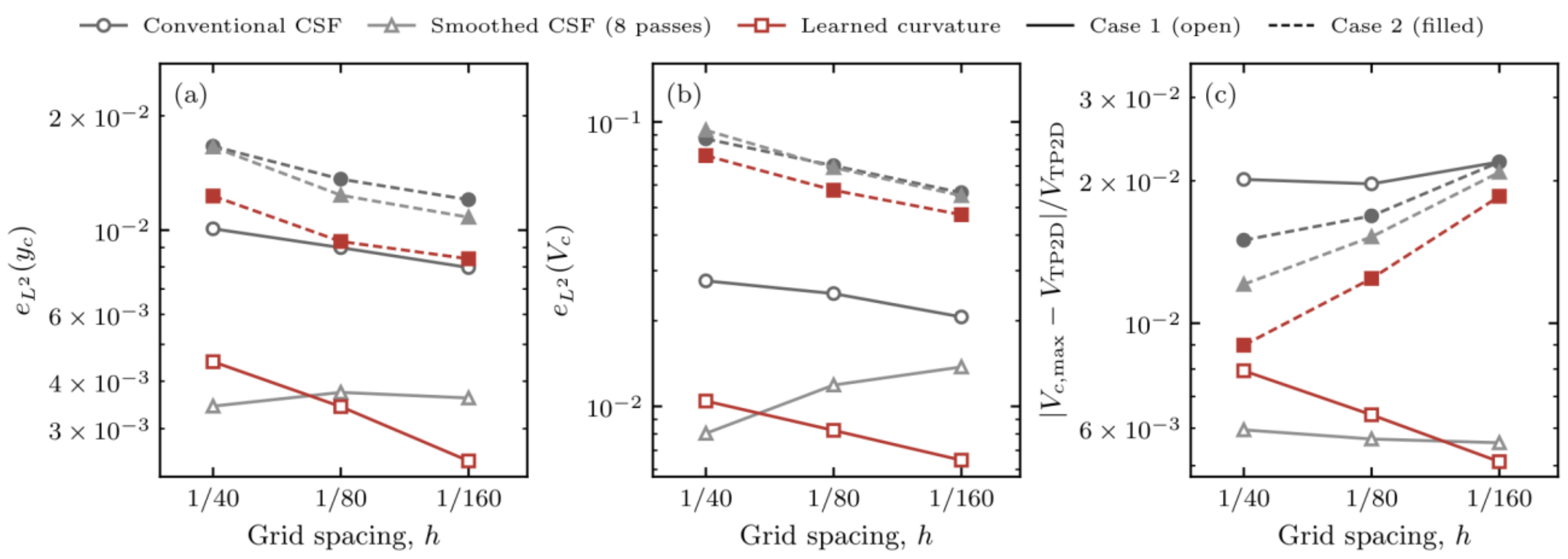

**Figure 12.** Grid dependence is reported through (a) centroid-height and (b) rise-velocity relative $L^2$ errors against digitised histories, and (c) first-peak deviations from TP2D values. Open symbols and solid lines denote case 1, while filled symbols and dashed lines denote case 2 for each curvature model.

**Table 2.** Benchmark scalars for three curvature models on three grids are compared with TP2D values [24], while integrated errors use the digitised reference histories. Dimensional quantities follow Section 5; integrated errors are percentages, and the sixth column gives circularity minima for case 1 and second velocity maxima for case 2. Simulation peak times use the 0.01 output resolution, while TP2D times retain their published precision; all listed configurations are excluded from curvature-network training.

| **Case** | **Model** | $h$ | $V_{c,\max}$ **(time)** | $y_c(3)$ | $c_{\min}$ **(time) or** $V_{c,\max,2}$ **(time)** | $e_{L^2}(y_c)$ | $e_{L^2}(V_c)$ |
|---|---|---|---|---|---|---|---|
| 1 | Conventional | 1/40 | 0.2368 (0.92) | 1.0654 | 0.8931 (1.93) | 1.01 | 2.76 |
| 1 | Conventional | 1/80 | 0.2369 (0.91) | 1.0670 | 0.8940 (1.95) | 0.90 | 2.49 |
| 1 | Conventional | 1/160 | 0.2364 (0.92) | 1.0703 | 0.8967 (1.90) | 0.80 | 2.06 |
| 1 | Smoothed | 1/40 | 0.2403 (0.93) | 1.0774 | 0.8918 (1.92) | 0.34 | 0.80 |
| 1 | Smoothed | 1/80 | 0.2403 (0.92) | 1.0750 | 0.8965 (1.97) | 0.37 | 1.19 |
| 1 | Smoothed | 1/160 | 0.2403 (0.92) | 1.0744 | 0.8971 (1.92) | 0.36 | 1.38 |
| 1 | Learned | 1/40 | 0.2398 (0.94) | 1.0776 | 0.8986 (1.91) | 0.45 | 1.04 |
| 1 | Learned | 1/80 | 0.2402 (0.93) | 1.0776 | 0.8982 (1.94) | 0.34 | 0.82 |
| 1 | Learned | 1/160 | 0.2405 (0.92) | 1.0785 | 0.8986 (1.93) | 0.25 | 0.65 |
| 1 | TP2D | – | 0.2417 (0.9213) | 1.0813 | 0.9013 (1.9041) | – | – |
| 2 | Conventional | 1/40 | 0.2486 (0.72) | 1.1034 | 0.2210 (1.82) | 1.66 | 8.72 |
| 2 | Conventional | 1/80 | 0.2481 (0.73) | 1.1098 | 0.2274 (1.92) | 1.36 | 7.02 |
| 2 | Conventional | 1/160 | 0.2469 (0.73) | 1.1147 | 0.2298 (1.99) | 1.20 | 5.64 |
| 2 | Smoothed | 1/40 | 0.2494 (0.74) | 1.1027 | 0.2219 (1.67) | 1.66 | 9.35 |
| 2 | Smoothed | 1/80 | 0.2486 (0.73) | 1.1117 | 0.2280 (1.89) | 1.24 | 6.91 |
| 2 | Smoothed | 1/160 | 0.2471 (0.73) | 1.1163 | 0.2298 (1.96) | 1.08 | 5.50 |
| 2 | Learned | 1/40 | 0.2501 (0.74) | 1.1112 | 0.2239 (1.73) | 1.23 | 7.61 |
| 2 | Learned | 1/80 | 0.2493 (0.74) | 1.1177 | 0.2299 (1.90) | 0.93 | 5.75 |
| 2 | Learned | 1/160 | 0.2477 (0.74) | 1.1206 | 0.2314 (1.97) | 0.84 | 4.72 |

| Case | Model | $h$ | $V_{c,\max}$ (time) | $y_c(3)$ | $c_{\min}$ (time) or $V_{c,\max,2}$ (time) | $e_{L^2}(y_c)$ | $e_{L^2}(V_c)$ |
|---|---|---|---|---|---|---|---|
| 2 | TP2D | – | 0.2524 (0.7332) | 1.1380 | 0.2434 (2.0705) | – | – |

## 6.4 Momentum-closure response and computational cost

The interface-gated model of Section 2.4 provides a separate momentum-transport assessment through the pipeline of Section 2.5, with learned curvature retained in the source simulations. Nine source simulations run at $h$=1/80 to $t$=2, sampling the capillary discretisation used during deployment rather than a different curvature formulation. Six cover density ratios 5.6–20, viscosity ratios 5.6–16.7, radii 0.19–0.24, and surface tensions 18–32, within the moderate-contrast family near case 1. The other three use density ratios 500, 667, and 1250, viscosity ratios 67–125, and surface tensions 2.2–3.2, sampling a high-contrast family relevant to case 2. Filtering with $f$=2 gives $h$=1/40 fields, and extraction at every fifth stored snapshot produces 27 684 samples in nine physical-parameter groups. One moderate-contrast group is reserved for validation and two groups, one from each family, for testing, so the partitions remain separate at the trajectory level. Within this dataset, the least-squares coefficient of Eq. (16) is negative in 52 % of samples, so the dissipative projection assigns those local backscatter targets to zero. The 7–16–1 network, trained for 5 000 Adam steps, gives normalised mean-squared errors of 0.044, 0.050, and 0.050 on training, validation, and test data. Its held-out coefficient of determination is $R^2$=0.47; Fig. 13(a) shows an increasing conditional prediction mean that falls below the diagonal above targets of 0.2. The compact seven-feature mapping therefore captures a substantial target trend while retaining regression towards the mean, with target populations at $y$=0 and $y$=1 reflecting projection and clipping.

The deployed field in Fig. 13(b), shown for case 2 at $t$=1 and $h$=1/80, occupies a one-to-two-cell band with maxima near the skirt's rear corners. At this snapshot, the closure-to-molecular kinematic-viscosity ratio remains below $8\times10^{-4}$, so the added dissipation remains localised where interfacial strain is largest. Over complete runs, peak closure viscosities are $1.0\times10^{-4}$ and $1.6\times10^{-5}$ for case 1, and $1.5\times10^{-4}$ and $6.5\times10^{-5}$ for case 2. These values correspond to the two coarser grids; the cap $b_\nu \nu$ remains inactive, and the contribution decreases with refinement, consistent with Proposition 1. Corresponding runs with and without the closure isolate the normalised rise-velocity changes in Fig. 13(c), with learned curvature common to both calculations. The largest changes are $1.7\times10^{-5}$ and $2.0\times10^{-6}$ in case 1, and $2.6\times10^{-4}$ and $8.8\times10^{-5}$ in case 2. Changes are negative at more than 90 % of output times, consistent with the dissipative form of Eq. (14), while integrated velocity errors change by at most 0.013 percentage points. The represented viscosity is approximately two to four orders below the molecular contribution in these comparisons, consistent with its limited influence on the laminar Re=35 dynamics.

The deployment guard also identifies a specific distribution difference: approximately 17–18 % of interface cells in the $h$=1/40 runs lie outside the training envelope. Offline checks identify $\Delta|\nabla\alpha|$ as the only feature triggering exclusions, because the compressed coarse-grid interface is sharper than the box-filtered fine-grid profile used for the targets. This observation supports matching filtered targets to the coarse solver's actual states, as emphasised in studies of training–deployment consistency [41, 62, 67, 71]. For the present bounded dissipative closure, the measured response is much smaller than that of curvature correction, so it serves as a controlled momentum-transport comparison.

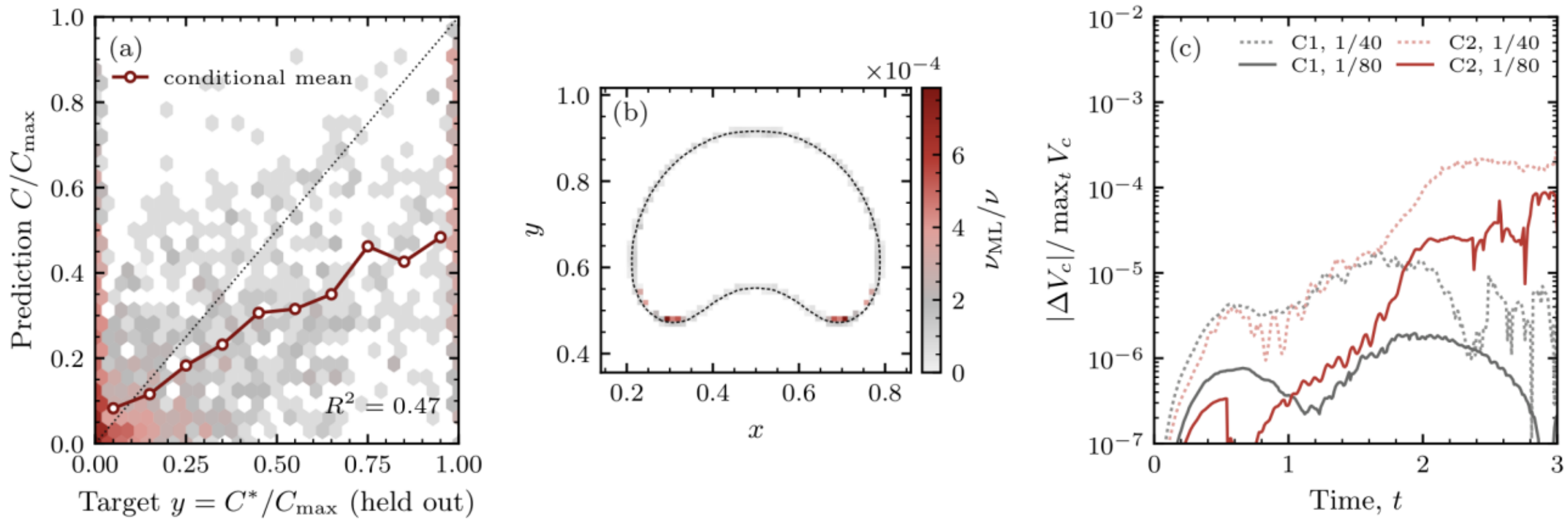

**Figure 13.** The interface-gated closure is assessed against Favre-filtered targets and through its deployed viscosity field and normalised influence on rise velocity. Panel (a) gives predictions against targets for two held-out groups through logarithmic hexagonal bin counts, with red denoting the most populated bins; a conditional prediction mean is superimposed in target bins. Panel (b) shows $\nu_{\rm ML}/\nu$ around case 2 at $t$=1 and $h$=1/80, with zero viscosity white and the interface dashed. Panel (c) gives closure-induced rise-velocity changes normalised by the maximum rise velocity for both benchmarks on the two coarser grids.

To complement the accuracy assessment, computational cost is measured by solver-reported processor time per step on the shared two-core workstation with varying concurrent workloads. Conventional costs are 16, 75–87, and 455–584 ms on successive grids. Learned curvature increases these costs to 48–59, 138–177, and 604–800 ms under the same grid sequence. The reported overhead factors are 3.0–3.8, 1.9–2.0, and 1.33–1.37, so the relative inference cost decreases as the mesh is refined. Each band cell requires 16 evaluations of the 25–128–128–1 network, which account for eight group elements and two phase-exchange signs per time step. Dense-layer products account for approximately $3\times10^5$ multiply–accumulate operations per band cell, with activation, indexing, and averaging costs additional to this count. Band work grows approximately as $h^{-1}$ in two dimensions, while bulk pressure-solution work grows more rapidly; the relative contribution of network evaluation therefore decreases with refinement. For comparison, smoothed curvature increases conventional cost by 10–15 %, and the momentum closure adds 4–8 % to the learned-curvature solver under the reported settings. The common capillary restriction of Eq. (27) remains an important contributor to total cost as refinement increases the required number of time steps. For completed benchmark runs, reported conventional-$h$=1/160 to learned-$h$=1/80 processor-cost ratios are 8.5 and 5.2 for cases 1 and 2, respectively. The learned intermediate grid improves the corresponding integrated metrics except case-2 velocity error, where 5.75 % compares with the conventional fine-grid value 5.64 %. These processor-cost ratios characterise the tested shared two-core workstation and its varying concurrent workloads; they remain specific to the reported platform and run conditions. The implementation and reported benchmarks concern serial execution on uniform, single-layer Cartesian meshes with two-dimensional velocity and interface variation. This setting defines the scope of the demonstrated accuracy and measured interface-work scaling across the three grids.

## 7 Conclusions

We developed a solver-consistent learned-curvature formulation for algebraic volume-of-fluid simulations in PGMLBUBBLEFOAM that preserves phase-exchange antisymmetry and Cartesian-grid invariance by construction. Training on solver-generated stencils and iterative data aggregation addressed feedback between capillary forcing and interface profiles and resolved the observed first-generation instabilities in the sampled periodic cases. The measured 11–22 % force-weighted curvature underestimate for radii of 2.5–40 cells quantitatively accounts for the conventional 14–17 % static pressure deficit in the examined configurations.

Across three grids, learned curvature reduces static pressure-jump errors to 2–6 % and time-averaged maximum capillary numbers by factors of 2.8–16. The time-integrated rise-velocity error decreases from 2.06–2.76 % to 0.65–1.04 % in the first Hysing benchmark and from 5.64–8.72 % to 4.72–7.61 % in the second. Tuned smoothed CSF remains more accurate for the static circle, while learned curvature gives smaller rise-velocity errors on the finer grids and consistent improvements over conventional curvature in both benchmarks.

The bounded momentum closure achieves a held-out $R^2$=0.47 but changes rise velocity by less than $3\times10^{-4}$ of its peak value in the tested configurations. This comparison identifies capillary-force discretisation as the more influential correction target for these laminar bubbles and supports training on the states encountered during solver deployment.

## Acknowledgements

The authors gratefully acknowledge Embry-Riddle Aeronautical University for providing access to the Vega High-Performance Computing Cluster and the computational resources used in this study.

## Declarations

### Conflict of interest

The authors declare that they have no competing financial or non-financial interests relevant to the research and results presented in this manuscript.

### Data and code availability

The solver source code, trained models, supporting datasets, benchmark configurations, and analysis scripts will be made publicly available in a GitHub repository after publication.